\documentclass{sig-alternate-10pt}

\usepackage[
      pass,
  ]{geometry}

\usepackage{hyperref}

\usepackage{graphicx}
\usepackage{amsmath}
\usepackage{mathtools}
\usepackage{url}
\usepackage{multirow}
\usepackage[caption=false]{subfig}
\usepackage{floatrow}
\usepackage{xcolor}
\usepackage{booktabs}
\usepackage{arydshln}
\usepackage{bm}
\usepackage[super]{nth}

\hypersetup{
    unicode=false,          
    pdftoolbar=true,        
    pdfmenubar=true,        
    pdffitwindow=false,     
    pdfstartview={FitH},    
    pdftitle={Pinpointing Delay and Forwarding Anomalies Using Large-Scale Traceroute Measurements}, 
    pdfauthor={Fontugne, et al.},     
    pdfsubject={Network Measurements},   
    pdfcreator={R. Fontugne},   
    pdfproducer={R. Fontugne}, 
    pdfkeywords={network disruption} {Internet} {latency} {traceroute}, 
    pdfnewwindow=true,      
    colorlinks=true,       
    linkcolor=blue,          
    citecolor=blue,        
    filecolor=magenta,      
    urlcolor=blue           
}

\newcommand{\TFIDF}[1]{}   
\newcommand{\arxiv}[1]{#1}   
\newcommand{\imc}[1]{} 

\DeclareMathOperator{\median}{median}

\DeclareMathOperator{\mad}{MAD}
\DeclareMathOperator{\tfidf}{tfidf}

\begin{document}

\title{Pinpointing Delay and Forwarding Anomalies Using Large-Scale Traceroute Measurements}

\numberofauthors{1}

\author{
    Romain Fontugne\\
       \affaddr{IIJ Research Lab}
\and
Emile Aben\\
       \affaddr{RIPE NCC}
\and 
Cristel Pelsser\\
       \affaddr{University of Strasbourg / CNRS}
\and 
Randy Bush\\
       \affaddr{IIJ Research Lab}
}

\maketitle
\begin{abstract}
Understanding data plane health is essential to improving Internet reliability and usability.  For instance, detecting disruptions in peer and
provider networks can identify repairable connectivity
problems. Currently this task is time consuming as it involves
a fair amount of manual observation, as an operator has poor visibility 
beyond their network's border.
  In this paper we leverage existing public RIPE Atlas
measurement data to monitor and analyze network conditions; creating no
new measurements.
We demonstrate a set of
complementary methods to detect network disruptions using traceroute 
measurements, and to report problems in near real time. A novel method of detecting changes in delay is used to identify
congested links, and a packet forwarding model is employed to predict
traffic paths and to identify faulty routers and links in cases of packet loss. 
In addition, aggregating results from each method allows us to easily monitor a
network and correlate related reports of significant
network disruptions, reducing uninteresting alarms.
Our contributions consist of a statistical approach to providing robust estimation
of Internet delays and the study of hundreds of thousands link delays.
We present three cases demonstrating that the proposed methods detect 
real disruptions and provide valuable insights, as well as surprising findings, 
on the location and impact of the identified events.
\end{abstract}

\section{Introduction}
The Internet's decentralized design allows disparate networks to cooperate and provides resilience to failure. 
However, significant network disruptions inevitably degrade users' connectivity. 
The first step to improve reliability is to understand the health of 
the current Internet. 
While network operators usually understand their own network's
condition, understanding the state of the multi-provider Internet beyond their own
network border remains a crucial but hard task.
Monitoring multiple networks' health is  difficult, and far too
often requires many manual observations.
For example, network operators' group mailing lists are a common way to signal 
and share knowledge about network disruptions \cite{banerjee:pam15}.
Manual network measurements, such as ping and traceroute assist in diagnosing connectivity
issues from a few vantage points but they suffer from poor visibility.

We investigate the potential of existing data from a large-scale measurement platform,
RIPE Atlas \cite{ripe:atlas}, to systematically detect and locate network disruptions.
The widespread deployment of Atlas probes provides an extensive view of the Internet
that has  proved beneficial for postmortem reports \cite{emile:nanog13,emile:amsix15,manojlovic:sanog16}.
Designing automated detection tools for such large-scale platforms is challenging.
The high variability of network performance metrics, such as round trip
time (RTT), is a key obstacle for reliable event detection \cite{pelsser:imc13}.
Beyond detecting network disruptions, pinpointing their location
is quite challenging due to traffic asymmetry and packet loss.

We examine these challenges (\autoref{sec:challenges}) and propose  methods to monitor 
the health of the vast number of networks probed by Atlas traceroutes.
First, we devise a method to monitor RTT from traceroute results and report 
links with unusual delays (\autoref{sec:delayAnalysis}).
This method takes advantage of the wide deployment of Atlas
by monitoring links from numerous vantage points, accurately measuring delay changes.
Second, we explore a packet forwarding model to learn and predict forwarding behavior
and pinpoint faulty routers experiencing sudden packet loss (\autoref{sec:forwardingAnomalies}).
Finally, we present a technique to aggregate these signals per network
and detect inter-related events (\autoref{sec:networkdisruptions}). 
These methods are all based on robust statistics which cope with outliers 
commonly found in traceroute measurements. 

The contributions of this work reside in the statistical approach to monitoring 
Internet delays.
Despite noisy RTT measurements, the introduced delay estimator infers very 
stable link delays and permits accurate predictions for anomaly detection.
It also enables the monitoring of delays and forwarding patterns for hundreds of 
thousands links.  
We provide our tools \cite{ihr:code} and report problems in near real time \cite{ihr:website, ihr:api}
so that other can build upon our work. 
Our proposal employs only existing data hence adding no burden to the network. 


To validate our methods we investigate three significant network events 
in 2015 (\autoref{sec:results}), each demonstrating key
benefits of our techniques.
The first analyzes the impact of a DDoS infrastructure attack.
The second shows congestion in a tier-1 ISP caused by inadvertent
rerouting of significant traffic. 
And the last presents connectivity issues at an Internet Exchange due to 
a technical fault.

\section{Dataset}
\label{sec:dataset}
To monitor as many links in the meshy Internet as possible, we need a vast number of vantage points
collecting network performance data. 
With its impressive spread across the globe and almost 10,000 probes constantly connected,
RIPE Atlas is the best candidate.
Atlas performs, among others, two classes of repetitive measurements providing an 
extensive collection of traceroute data publicly available in near real time.
The first type, \emph{builtin} measurements, consists of traceroutes
from all Atlas probes to instances of the 13 DNS root servers every 30 minutes.
Due to the wide distribution of probes and the anycast DNS root server deployment, 
this is actually to over 500 root server instances.
The second type, \emph{anchoring} measurements, are traceroutes to 189 collaborative
servers (super probes) from about 400 normal probes every 15 minutes. 
All measurements employ Paris traceroute \cite{augustin:imc06} to mitigate issues
raised by load balancers and link aggregation \cite{pelsser:imc13}.

We have analyzed the builtin and anchoring measurements
from May \nth{1} to December \nth{31} 2015, corresponding to a total of 2.8 billion IPv4 traceroutes (1.2 billion IPv6 traceroutes)
from a total of 11,538 IPv4 probes (4,307 IPv6 probes) connected within the eight 
studied months.

As our study relies solely on traceroute results the scope and terminology
of this paper are constrained to the IP layer.
That is, a link refers to a pair of IP addresses rather than a physical
cable.

Consequently, the proposed methods suffer from common limitations faced by traceroute data
\cite{luckie:imc08,roughan:ieee11,luckie:imc14}.
Traceroute visibility is limited to the IP space, hence, changes at lower layers that
are not visible at the IP layer can be misinterpreted.
For example, the RIPE Atlas data reports MPLS information if routers support RFC4950.
But for routers not supporting RFC4950, the reconfiguration of an MPLS
tunnel is not visible with traceroutes while being likely to impact observed delays.
The RTT values reported by traceroute include both network delays and
routers' slow path delay \cite{luckie:imc14}.
Therefore, the delay changes found using traceroute data are not to be taken as 
actual delay increases experienced by TCP/UDP traffic, though they are
good for detecting network damage.

\section{Challenges and Related Work}
\label{sec:challenges}
Monitoring network performance with traceroute raises three key challenges. In 
this section, we present these challenges, discuss how they were tackled in 
previous work, and give hints of our approach to be discussed in detail later.

\noindent{\bf Challenge 1: Traffic asymmetry.} \
\label{sec:trafficAsymmetry}
Traceroutes are a rich source of information for monitoring Internet delay. 
They reveal the path to a destination and provide RTTs for every router 
on this path.
Each RTT value is the sum of the time spent to reach a certain IP address 
and the travel time for the corresponding reply.
Due to the asymmetry and diversity of routes \cite{renata:imc03,zheng:pam05} 
the paths taken by the forwarding and returning packets often differ; 
also traceroute is unable to reveal IP addresses on the return path.
Path asymmetry is very common; past studies report about 90\% 
of AS-level routes as asymmetric \cite{schwartz:infocom10,vries:aims15}.  
For these reasons one must take particular care when comparing RTT values for different hops.

\begin{figure}
\centering
    \vspace*{-3mm}
\subfloat[\scriptsize Round-trip to router B (blue) and C (red).\label{fig:tracerouteExampleC}\label{fig:tracerouteExampleB}]{\includegraphics[width=.45\columnwidth]{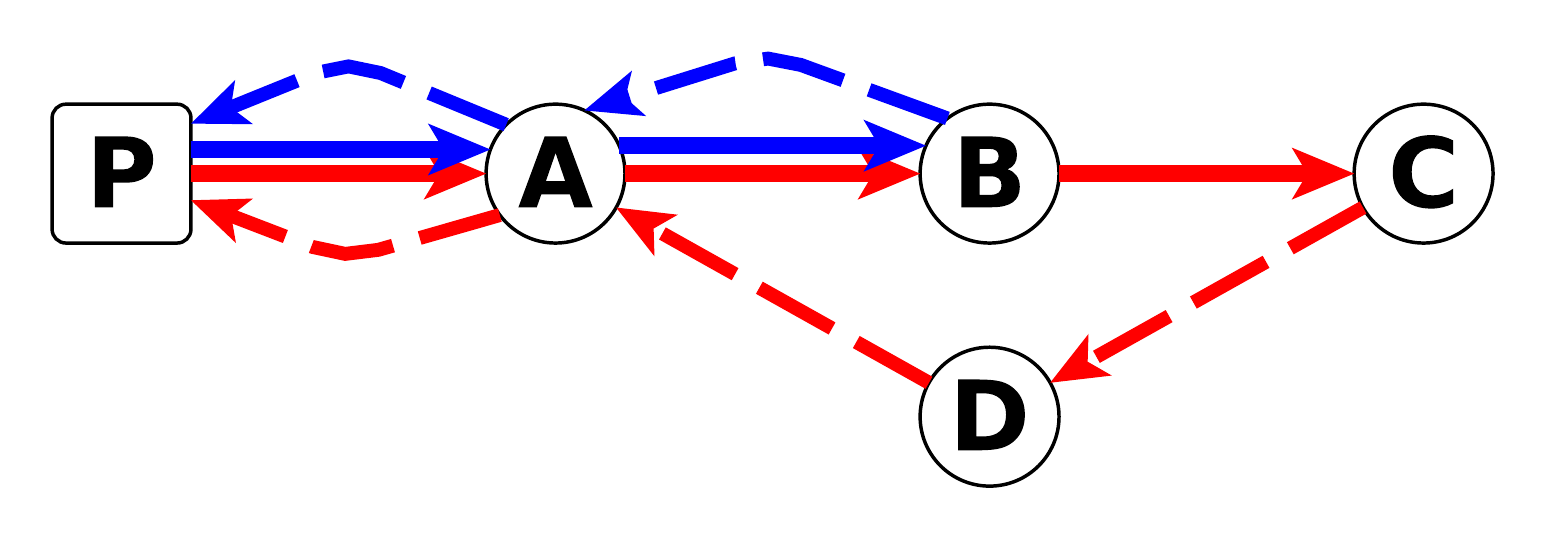}}
~~~~~~~
\subfloat[\scriptsize Difference of the two round-trips ($\Delta_{PBC}$).\label{fig:tracerouteExampleCB}]{\includegraphics[width=.32\columnwidth]{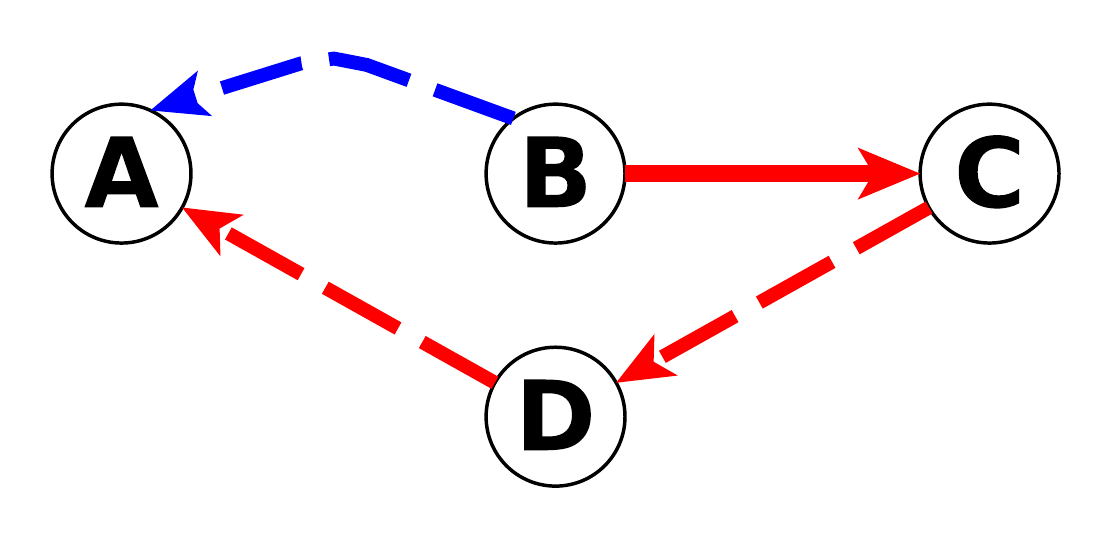}}
    \vspace*{-3mm}
\caption{Example of traceroute results with different return paths.
    $P$ is the probe initiating the traceroute. $A$, $B$, and $C$ are routers reported
    by traceroute. $D$ is a router on the return path, unseen in the traceroute. 
    Solid lines represent the forward paths, dashed the return paths.
}
\label{fig:tracerouteExample}
\end{figure}
For instance, quantifying the delay between two adjacent hops can be baffling.
Figure \ref{fig:tracerouteExample} illustrates this by breaking down
the RTT from the probe P to router B (blue in Fig.~\ref{fig:tracerouteExampleB}) and the one to the following 
hop, router C (red in Fig.~\ref{fig:tracerouteExampleC}).
The solid lines represent the forward path exposed by traceroute, and the dotted
the unrevealed return path.
If we want to measure the delay between routers B and C using only the information provided by traceroute (i.e. solid lines 
in Fig.~\ref{fig:tracerouteExample}), one is tempted to compute the delay between B 
and C as the difference between the RTT to B and the one to C.
But the resulting value is likely incorrect when forward and return paths 
are asymmetric.
Packets returning from 
C are not going through B but D, a router not seen on the forward path.
If one is monitoring the difference between the two RTTs
over time and identifying an abnormality, then it is unclear if a change 
is due to abnormal delay on link $BC$, $CD$, $DA$, or $BA$ (Fig.~\ref{fig:tracerouteExampleCB}). 

Previous studies approach this using reverse traceroute techniques
based on IP options to expose the return path \cite{katz:nsdi10,marchetta:pam14}.
Using these techniques Luckie et al. \cite{luckie:imc14} filter out routers with 
different forward and return paths and characterize congestion for the remaining routers.
Due to the limitations of these reverse traceroute techniques \cite{donato:pam12} and the strong asymmetry
of Internet traffic \cite{vries:aims15}, they could study only $29.9\%$ of the routers 
observed in their experiments.

Coordinated probing from both ends of the path is another way to reveal asymmetric
paths and corresponding delays \cite{deng:icnp08,chand:conext15}.
However, coordinated probing requires synchronized control on hosts located at both 
ends of the path, which is difficult in practice and limits the probing surface.

Tulip \cite{mahajan:sosp03} and cing \cite{cing:infocom03} bypass the traffic 
asymmetry problem by measuring delays with ICMP options but require routers to 
implement these options.

In Section \ref{sec:diffRtt} we review the asymmetric paths problem and propose a new
approach that takes advantage of multiple probes and path diversity to accurately monitor
delay fluctuations for links visited from different vantage points.

\noindent{\bf Challenge 2: RTT variability.} \
As packets traverse multiple links, routers, queues, and middleboxes,
they are exposed to multiple sources of delay that result in complex RTT dynamics.
This phenomenon has been studied since the early years of the Internet and
is still of interest as comprehensive understanding of delay
is a key step to understanding network conditions \cite{rizo:comlet14,romain:infocom15,singla:hotnets14,owen:imc15}.
Simply stated, monitoring delay is a delicate task because RTT samples
are contaminated by various noise sources.
In the literature, RTTs are monitored with different goals in mind. 
Minimum RTT values reveal propagation and transmission delays but filter out 
delays from transient congestion, so are commonly used to compute
geographic distance in IP geolocation systems \cite{katz:imc06,zhang:ton10}.
Studies focusing on queuing delays usually rely on RTT percentiles \cite{aikat:IMC03,markopoulou:comcom06};
there is however no convention to choose specific quantiles.
For instance, Chandrasekaran et al. \cite{chand:conext15} define the $10^{th}$ percentile as the \emph{baseline}
RTT and the $90^{th}$ percentile as \emph{spikes} (i.e. sudden RTT increases),
in the same study they also report results for the $5^{th}$ and $95^{th}$ percentiles.

We monitor the median RTT (i.e. $50^{th}$ percentile) which  
accounts for high delays only if they represent the majority of the RTT samples. 
Section \ref{sec:rttChangeDetect} presents the other robust statistics we employ to 
analyze RTT measurements.

\noindent{\bf Challenge 3: Packet loss.} \
Delay is an important but insufficient indicator to identify connectivity issues. 
In worst-case scenarios networks fail to transmit packets, and the lack of samples
clouds delay measurements.
Increases in delay and packet loss are not necessarily correlated \cite{markopoulou:comcom06}.
Congestion provides typical examples where both metrics are affected \cite{sommers:sigcomm05}, 
but routers implementing active queue management (e.g. Random Early Detection \cite{floyd:ton93})
can mitigate this \cite{luckie:imc14}, as the routers drop packets to avoid
significant delay increase.
Other examples include bursts of lost packets on routing failure \cite{wang:sigcomm06}. 
We stress that a comprehensive analysis of network conditions must track both
network delay and packet loss.

Packet loss is sometimes overlooked by congestion detection systems.
For instance, Pong \cite{deng:icnp08} and TSLP \cite{luckie:imc14} probe routers
to monitor queuing delays, but users are left with no guidance in the case of lost probes.
Consequently, studies using these techniques tend to ignore incomplete data
due to lost packets (e.g. 25\% of the dataset is disregarded in ref. \cite{chand:conext15}),
and potentially miss major events.

Detecting packet loss is of course an easy task; the key difficulty is to locate
where the packets are dropped.
Several approaches have been previously proposed to address this. 
The obvious technique is to continuously probe routers, or networks, 
and report packet loss or disconnections \cite{mahajan:sosp03,trinocular}.
This is, however, particularly greedy in terms of network resources, hence, difficult 
to deploy for long-term measurements. 
Another approach employs both passive and active monitoring techniques to build
end-to-end reference paths, passively detect packet loss, and actively locate path changes \cite{zhang:osdi04}.
Approaches using only passive measurements are also possible; although wide coverage requires
collection of flow statistics from many routers \cite{gu:infocom09}.

In Section \ref{sec:forwardingAnomalies} we introduce a forwarding anomaly detection 
method that complements the proposed RTT analysis method (\autoref{sec:delayAnalysis}).
It analyzes traceroute data and creates reference forwarding patterns for each router. 
These patterns are used to locate routers that drop packets in abnormal situations.

\imc{ Further comparisons with related works are provided in \cite{tartiflette:arxiv} Appendix A. }
\arxiv{ Further comparisons with related works are provided in Appendix \ref{sec:comparison}.}


\section{In-Network Delays}
\label{sec:delayAnalysis}
We now describe our approach to detecting abnormal delay changes
in wide-area traceroute measurements.
To address the traffic asymmetry challenge we propose monitoring a link's delay 
using Atlas probes from multiple ASs (\autoref{sec:diffRtt}).
Then, we use a robust detector to identify abnormal delay changes (\autoref{sec:rttChangeDetect}).

\subsection{Differential RTT}
\label{sec:diffRtt}
As stated in Section \ref{sec:trafficAsymmetry}, locating delay changes from 
traceroute data is challenging because of traffic asymmetry.
We address this challenge by taking advantage of the topographically-diverse 
deployment of Atlas probes.

Let's revisit the example of Figure \ref{fig:tracerouteExample} and introduce our notation.
$RTT_{PB}$ stands for the round-trip-time from the probe $P$ to the router $B$.
The difference between the RTT from $P$ to the two adjacent routers, $B$ and $C$,
is called differential RTT and noted $\Delta_{PBC}$.
The differential RTT of Figure \ref{fig:tracerouteExampleCB} is decomposed as follows:
\begin{align}
    \Delta_{PBC} & = RTT_{PC} - RTT_{PB} \\
                 & = \delta_{BC} + \delta_{CD} + \delta_{DA} - \delta_{BA} \\
                 & = \delta_{BC} + \varepsilon_{PBC} \label{eq:diffRtt}
\end{align}
where $\delta_{BC}$ is the delay for the link $BC$ and $\varepsilon_{PBC}$ is
the time difference between the two return paths.

$\Delta_{PBC}$ alone gives a poor indication of the delay of link BC because the two 
components, $\delta_{BC}$ and $\varepsilon_{PBC}$, are not dissociable.
Nonetheless, these two variables are independent and controlled by different factors. 
The value of $\delta_{BC}$ depends only on the states of routers $B$ and $C$, and is
unrelated to the monitoring probe $P$.
In contrast, $\varepsilon_{PBC}$ is intimately tied to $P$, the destination for 
the two return paths.

Assuming that we have a pool of $n$ probes $P_i, i\in[1,n]$, all with different 
return paths from $B$ and $C$;
then, the differential RTTs for all probes, $\Delta_{P_iBC}$, share the same $\delta_{BC}$ 
but have independent $\varepsilon_{P_iBC}$ values.
The independence of $\varepsilon_{P_iBC}$ also means that the distribution of 
$\Delta_{P_iBC}$ is expected to be stable over time if $\delta_{BC}$ is constant.
In contrast, significant changes in $\delta_{BC}$ influence all differential RTT
values and the distribution of $\Delta_{P_iBC}$  shifts along with the $\delta_{BC}$ changes.
Monitoring these shifts allows us to discard uncertainty from return paths ($\varepsilon_{P_iBC}$)
and focus only on delay changes for the observed link ($\delta_{BC}$). 

Now let's assume the opposite scenario  where $B$ always pushes returning packets to $A$,
the previous router on the forwarding path (see link $AB$ in Fig.~\ref{fig:tracerouteExample}). 
In this case $\varepsilon_{P}$ represents the delay between $B$ and $A$; hence, 
Equation \ref{eq:diffRtt} simplifies as:
\begin{equation}
    \Delta_{PAB} = \delta_{AB} + \delta_{BA}. 
\end{equation}
Meaning the differential RTT $\Delta_{PAB}$ stands for the delays between router
$A$ and $B$ in both directions.
This scenario is similar to the one handled by TSLP \cite{luckie:imc14}, and in 
the case of delay changes, determining which one of the two directions is affected requires 
extra measurements (see \cite{luckie:imc14} Section 3.4).

In both scenarios, monitoring the distribution of differential RTTs
detects delay changes between the adjacent routers.  Note
that we are looking exclusively at differential RTT fluctuations rather
than their absolute values.  The absolute values of differential RTTs
can be misleading; as they include error from return paths, they cannot
account for the actual link delay.  In our experiments we observe
negative differential RTTs, $\Delta_{PXY}<0$, meaning that $Y$ has a
lower RTT than $X$ due to traffic asymmetry (see Fig.~\ref{fig:kroot:decix0} and \ref{fig:kroot:stpe0}).

\subsection{Delay change detection}
\label{sec:rttChangeDetect}
The theoretical observations of the previous section are the fundamental 
mechanisms of our delay change detection system.
Namely, the system collects all traceroutes initiated in a 1-hour time bin and 
performs the following five steps:
(1)
 Compute the differential RTTs for each link (i.e. pair of adjacent IP addresses 
        observed in traceroutes).
(2)
 Links that are observed from only a few ASs are discarded.
(3)
 The differential RTT distributions of the remaining links are characterized
        with nonparametric statistics,
(4)
 and compared to previously computed references in order to identify abnormal delay changes.
(5)
 The references are updated with the latest differential RTT values.
The same steps are repeated to analyze the subsequent time bins.
The remainder of this section details steps for handling differential RTTs (i.e. steps 1, 3, 4, and 5).
Step 2 is a filtering process to discard links with ambiguous differential RTTs 
and is discussed later in Section \ref{sec:probeDiversity}.

\subsubsection{Differential RTT computation}
The first step is calculating the difference between RTT values 
measured for adjacent routers.
Let $X$ and $Y$ be two adjacent routers observed in a traceroute initiated by the
probe $P$.
The traceroute yields from one to three values for $RTT_{PX}$ and $RTT_{PY}$.
The differential RTT samples, $\Delta_{PXY}$ are computed for all possible combinations  
$RTT_{PY}-RTT_{PX}$; hence, we have from one to nine differential RTT samples per probe.
In the following, all differential RTTs obtained with every probe are denoted $\Delta_{XY}$,
or $\Delta$ when  confusion is not likely. 

\subsubsection{Differential RTTs characterization}
\label{sec:rttCharac}
This step characterizes the distributions of differential RTTs $\Delta_{XY}$ obtained in the
previous step, in order to compute a normal reference and detect significant deviations
from it.
In practice, these anomalies are detected using a variant of the Central Limit Theorem (CLT).
The original CLT states that, regardless the distribution of $\Delta_{XY}$,
its arithmetic mean is normally distributed if the number of samples is relatively large. 
If the underlying process changes, in our case the delays for $X$ and $Y$, 
then the resulting mean values deviate from the normal distribution 
and are detected as anomalous. 

Our preliminary experiments suggest that the frequent outlying values found in RTT 
measurements greatly affect the computed mean values; thus an impractical
number of samples is required for the CLT to hold.
To address this we replace the arithmetic mean by the median.
This variant of the CLT is much more robust to outlying values 
and requires less samples to converge to the 
normal distribution \cite{wilcox:stats10}.
\begin{figure}
    \includegraphics[width=\columnwidth]{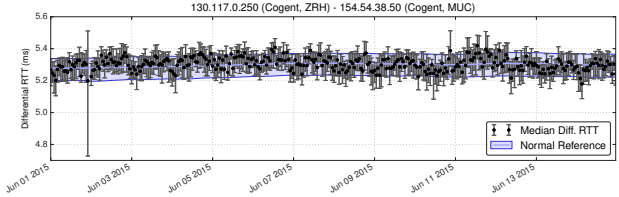}
    \caption{Example of median differential RTTs for a pair of IP addresses from Cogent Communications (AS174).
    Every median differential RTT is computed from a 1-hour time window, the error bars
    are the 95\% confidence intervals obtained by the Wilson Score and the 
    normal reference is derived from these intervals.}
    \label{fig:exampleDiffRtt} 
    \vspace*{-5mm}
\end{figure}
\begin{figure}
    \vspace*{-5mm}
    \subfloat[\vspace*{-1mm}\scriptsize Median diff. RTT. \label{fig:qqplot_median}]{
    \includegraphics[width=.45\columnwidth]{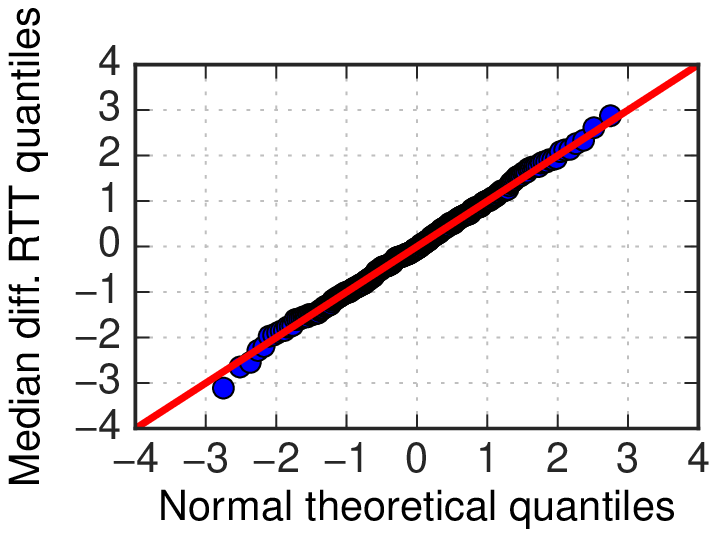}}
    \subfloat[\vspace*{-1mm}\scriptsize Mean diff. RTT. \label{fig:qqplot_mean}]{
    \includegraphics[width=.45\columnwidth]{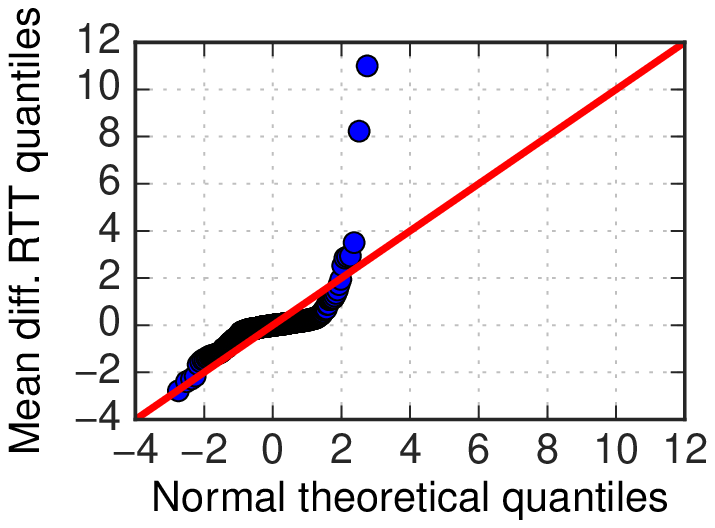}
    }
    \vspace*{-3mm}
    \caption{Normality tests for the same data as Figure \ref{fig:exampleDiffRtt}.
    Q-Q plots of the median and mean differential RTT versus a normal distribution.} 
\end{figure}
Figure \ref{fig:exampleDiffRtt} depicts the hourly median differential RTTs
(black dots) obtained for a pair of IP addresses from Cogent networks (AS174) during
two weeks in June 2015.
This link is observed by 95 different probes between June \nth{1} and June \nth{15}.
The raw differential RTT values exhibit large fluctuations; the standard deviation ($\sigma=12.2$)
is almost three times larger than the average value ($\mu=4.8$).
Despite this variability, the median differential RTT is remarkably
steady, all values lie between 5.2 and 5.4 milliseconds (Fig.~\ref{fig:exampleDiffRtt}).
Significant fluctuations of the median would indicate trustworthy delay changes on that link.


We confirm that the employed CLT variant holds very well with differential RTTs.
Figure \ref{fig:qqplot_median} compares the quantiles of the computed medians to 
those of a normal distribution.
As all points are in line with the $x=y$ diagonal, the computed median differential RTTs 
fit a normal distribution quite well.
In contrast, the mean differential RTT is not normally distributed (Fig.~\ref{fig:qqplot_mean}).
By manually inspecting the raw RTT values, we found 125 outlying values 
(i.e. greater than $\mu+3\sigma$) that greatly alter the mean. 
These outliers are isolated events spread throughout the two weeks, and are attributed 
to measurement errors.
Despite the large number of probing packets going through this link, the mean differential
RTTs are greatly altered by these few outliers. 
These observations support our choice for the median CLT variant against
the original CLT.

To account for uncertainty in the computed medians, we also calculate 
confidence intervals.
In the case of the median, confidence intervals are usually formulated as a 
binomial calculation and are distribution free \cite{gibbons:nonparametric11}.
In this work we approximate this calculation with the Wilson score \cite{wilson:score27}
since it has been reported to perform well even with a small number of samples \cite{newcombe:confint98}.
The Wilson score is defined as follows:
\begin{equation}
    w = \frac{1}{1+\frac{1}{n}z^2} \left( p + \frac{1}{2n}z^2 \pm z \sqrt{ \frac{1}{n}p(1-p)+\frac{1}{4n^2}z^2} \right)
    \label{eq:wilson}
\end{equation}
where $n$ is the number of samples, the probability of success $p$ is set to 0.5
in the case of the median, and $z$ is set to 1.96 for a 95\% confidence level.
The Wilson score provides two values, hereafter called $w_l$ and $w_u$, ranging in $[0, 1]$.
Multiplying $w_l$ and $w_u$ by the number of samples gives the rank of the lower 
and upper bound of the confidence interval, namely $l=nw_l$ and $u=nw_u$.

For example, let $\Delta^{(1)},...,\Delta^{(n)}$ be the $n$ differential 
RTT values obtained for a single link, and assume these values are ordered, i.e.
$\Delta^{(1)}\leq\Delta^{(2)}\leq...\leq\Delta^{(n)}$.
Then, for these measures the lower and upper bound of the confidence interval are given 
by $\Delta^{(l)}$ and $\Delta^{(u)}$. 

Based solely on order statistics, the Wilson score produces asymmetric confidence 
intervals in the case of skewed distributions, which are common for RTT distributions \cite{romain:infocom15}.
Further, unlike a simple confidence interval based on the standard deviation, 
this non-parametric technique takes advantage of order statistics to discard 
undesirable outliers.

The whiskers in Figure \ref{fig:exampleDiffRtt} depict the confidence intervals 
obtained for the Cogent link discussed above.
These intervals are consistent over time and show that the median differential
RTT for this link reliably falls between 5.1 and 5.5 milliseconds.
The large confidence interval reported on June 1$st$ illustrates an example 
where RTT measures are noisier than other days; yet we stress that the median value 
and confidence interval are compatible with those obtained by other time bins.
The following section describes how we identify statistically deviating differential RTTs.

\subsubsection{Anomalous delays detection}
\label{sec:delayAnomaly}
A delay change results in a differential RTT distribution shift; therefore a significant change 
in the corresponding median differential RTT value.
Assume we have a reference median and its corresponding 95\% confidence interval
that represents the usual delay measured for a certain link
(the calculation of such reference is addressed in \autoref{sec:normalRef}).
To measure if the difference between an observed median and the reference is  
statistically significant we examine the overlap between their confidence 
intervals.
If the two confidence intervals are not overlapping, we conclude that there is a
statistically significant difference between the two medians \cite{schenker:cioverlap}
so we report the observed median as anomalous.
As a rule of thumb we discard anomalies where the difference between the two medians
is lower than 1ms (in our experiments these account for 3\% of the reported links). 
Although statistically meaningful, these small anomalies are less relevant for 
the study of network disruption.

The deviation from the normal reference is given by the gap between the 
two confidence intervals.
Let $\bar\Delta^{(l)}$ and $\bar\Delta^{(u)}$ be, respectively, the lower and
upper bound of the reference confidence interval and $\bar\Delta^{(m)}$ the 
reference median. 
Then, the deviation from the normal reference of the observed differential RTTs, $\Delta$, is defined as:
\begin{equation}
d(\Delta) = 
   \begin{dcases}
       \frac{\Delta^{(l)} - \bar\Delta^{(u)}}{ \bar\Delta^{(u)} - \bar\Delta^{(m)}}, & \text{if } \bar\Delta^{(u)} < \Delta^{(l)} \\
        \frac{\bar\Delta^{(l)} - \Delta^{(u)}}{\bar\Delta^{(m)} - \bar\Delta^{(l)}}, & \text{if } \bar\Delta^{(l)} > \Delta^{(u)} \\
       ~0, & \text{otherwise.}
   \end{dcases}
   \label{eq:delay:deviation} 
\end{equation}

This deviation represents the gap separating the two confidence intervals 
and is relative to the usual uncertainty measured by the reference confidence
interval. 
Values close to zero represent small delay changes while large values represent
important changes.

Figure \ref{fig:exampleDiffRtt} exhibits confidence intervals along with
the corresponding normal reference.
As the reference intersects with all confidence intervals, no anomaly 
is reported for this link.
The evaluation section presents several examples of anomalies. 
For example, Figure \ref{fig:kroot:decix0} depicts two confidence intervals deviating
from the normal reference on November \nth{30}.

\subsubsection{Normal reference computation}
\label{sec:normalRef}
In the previous section we assumed a reference differential RTT 
distribution for each link.  We will now show how to compute this.
The goal of the references is to characterize the usual delays of observed links.
As median differential RTT values are normally distributed (\autoref{sec:rttCharac}),
the expected median value for a link is simply obtained as the arithmetic 
mean of previously observed medians for that link.
Because anomalies might impair mean values and make them irrelevant as references,
we employ exponential smoothing to estimate the medians' mean value to
reduce the impact of anomalies.
Exponential smoothing also facilitates the online implementation of our delay change method
for near real time analysis \cite{ihr:code, ihr:website}.
Let $m_{t} = \Delta^{(m)}$ be the median differential RTT observed for a certain
link in time bin $t$, and, $\bar m_{t-1} = \bar \Delta^{(m)}$ be the reference 
median computed with median differential RTTs observed in the previous time bin, $t-1$.
Then the next reference median, $\bar m_{t}$ is defined as:
\begin{equation}
    \bar m_{t} = \alpha m_{t} + (1-\alpha) \bar m_{t-1}
    \label{eq:expSmoothing}
\end{equation}
The only parameter for the exponential smoothing, $\alpha\in(0,1)$, controls the importance
of new measures as opposed to the previously observed ones.
In our case a small $\alpha$ value is preferable as it lets us mitigate the 
impact of anomalous values.
The initial value of the reference, $\bar m_{0}$, is quite
important when $\alpha$ is small.
We arbitrarily set this value using the first three time bins,
namely, $\bar m_{0} = \median(m_1, m_2, m_3)$. 

For the reference confidence interval, the lower  and upper
bounds (resp. $\bar\Delta^{(l)}$ and $\bar\Delta^{(u)})$ are computed in the 
same way as the reference median ($\bar\Delta^{(m)}$) but using the boundary
values given by the Wilson score (i.e. $\Delta^{(l)}$ and $\Delta^{(u)}$).

\subsection{Probe diversity}
\label{sec:probeDiversity}
The above differential RTT analysis applies only under certain conditions.
Section \ref{sec:diffRtt} shows that monitoring $\Delta_{XY}$ 
reveals delay changes between router $X$ and $Y$ only if the following
hold true.  (1) The link is monitored by several probes and the return paths to 
these probes are disparate.
(2) All returning packets are also going through the link $XY$ but in the opposite direction.
Therefore, if we have differential RTT values $\Delta_{XY}$ from ten probes 
which share the same asymmetric return path, we cannot distinguish delay changes on $XY$ from delay changes in 
the return path, so these differential RTT values cannot be used.

To filter out ambiguous differential RTTs
we avoid links monitored 
only by probes from the same AS (thus more likely to share the same return path due to 
common inter-domain routing policies);
but instead, take advantage of the wide deployment of Atlas probes and focus
on links monitored from a variety of ASs.
We devise two criteria to control the diversity of probes monitoring a link.

The first criterion filters out links that are monitored by probes from 
less than 3 different ASs.
The value 3 is empirically set to provide conservative results and can be lowered to increase the number of monitored links but at the cost of result accuracy. 
To determine this value we make the following hypothesis.
Links where the error added by return paths is not mitigated by probe diversity are reported more frequently as their 
differential RTTs also account for links on the return path.
For links visited by probes from at least 3 different ASs we observe a weak positive correlation  (0.24) between the average number of reported alarms and the number of probes monitoring a link.
Meaning that links observed by a small number of diverse probes are not reported more than those monitored by a large number of probes, thus a small diversity of return paths is enough to mitigate the error added by return paths. 
 

This simple criterion allows us to avoid ambiguous results when links are monitored 
from only a few ASs, but is insufficient to control probe diversity.
For instance, a link $XY$ is monitored by 100 probes located in 5 different ASs
but 90 of these probes are in the same AS.
Then, the corresponding differential RTT distribution is governed by the return
path shared by these 90 probes, meaning that delay changes on this return path
are indistinguishable from delay changes on $XY$.

The second criterion finds links with an unbalanced number of probes per AS.
Measuring such information dispersion is commonly addressed using normalized entropy.
Let $A= \left\{a_i | i \in [1,n] \right\}$ be the number of probes for each of the $n$ ASs monitoring 
a certain link, then the entropy $H(A)$ is defined as:
    $H(A) = -\frac{1}{\ln n}\sum_{i=1}^{n} P(a_i) \ln P(a_i)$.
Low entropy values, $H(A)\simeq0$, mean that most of the probes are 
concentrated in one AS, and, high entropy values, $H(A)\simeq1$, indicate 
that probes are evenly dispersed among ASs.
This second criterion ensures that analyzed links feature an entropy 
$H(A)>0.5$.

If the second criterion is not met (i.e. $H(A)\leq0.5$) the link is not discarded.
Instead, a probe from the most represented AS (namely AS $i$ such as $a_i = \max(A)$) 
is randomly selected and discarded, thus increasing the value of $H(A)$.
This process is repeated until $H(A)>0.5$, hence the corresponding differential 
RTTs are relevant for our analysis.

    \subsection{Theoretical limitations}
In our experiments we conservatively set the time bin to one hour, consequently, 
the shortest event we can detect for a link monitored by three vantage points is 33 minutes 
long 
\imc{(see \cite{tartiflette:arxiv} Appendix B).}
\arxiv{(see Appendix \ref{appendix:theo_limitations}).}
Using measurements with a high probing rate overcomes this limitation, for
instance, anchoring measurements can detect events lasting only 9 minutes.

Low frequency traceroute measurements originally designed for topology discovery are 
not suitable for our approach. For example, the \emph{IPv4 Routed /24 Topology Dataset} from 
CAIDA \cite{caida:topology} has a 48 hour cycle which is not appropriate to monitor transient 
delay changes.

\section{Forwarding Anomalies}
\label{sec:forwardingAnomalies}
Latency is a good indicator of network health, but deficient in certain cases.
For example, if traffic is rerouted or probing packets are lost then the lack of
RTT samples impedes delay analysis.
We refer to these cases as forwarding anomalies.
In this section we introduce a method to detect forwarding anomalies, complementing
the delay analysis method presented in Section~\ref{sec:delayAnalysis}.

A forwarding anomaly can be legitimate, for example rerouted traffic,
but it can also highlight compelling events such as link failures or routers 
dropping packets.
Using traceroute data, such events appear as router hops vanishing from our dataset.
So our approach monitors where packets are forwarded 
and constructs a simple packet forwarding model 
(\autoref{sec:forwardingModel}).
This model allows us to predict next hop IP addresses in traceroutes, thus 
detecting and identifying disappearing routers (\autoref{sec:forwardingAnomaly}).

\subsection{Packet forwarding model}
\label{sec:forwardingModel}
The proposed packet forwarding model learns the next hops
usually observed after each router from past traceroute data.
Because routers determine next hops based on the packet destination IP address,
we compute a different model for each traceroute target.

Let us consider traceroutes from all probes to a single destination in the same time bin.
For each router in these traceroutes we record the adjacent nodes to
which packets have been forwarded.
We distinguish two types of next hop, responsive and unresponsive ones.
The responsive next hops are visible in traceroutes as they send back ICMP messages 
when a packet TTL expires. 
Next hops that do not send back ICMP packets to the probes or drop packets 
are said to be unresponsive and are indissociable in traceroutes.

\begin{figure}
    \vspace*{-2mm}
    \subfloat[\scriptsize Usual forwarding pattern.\label{fig:forwardingExample0}]{~\includegraphics[width=.28\columnwidth]{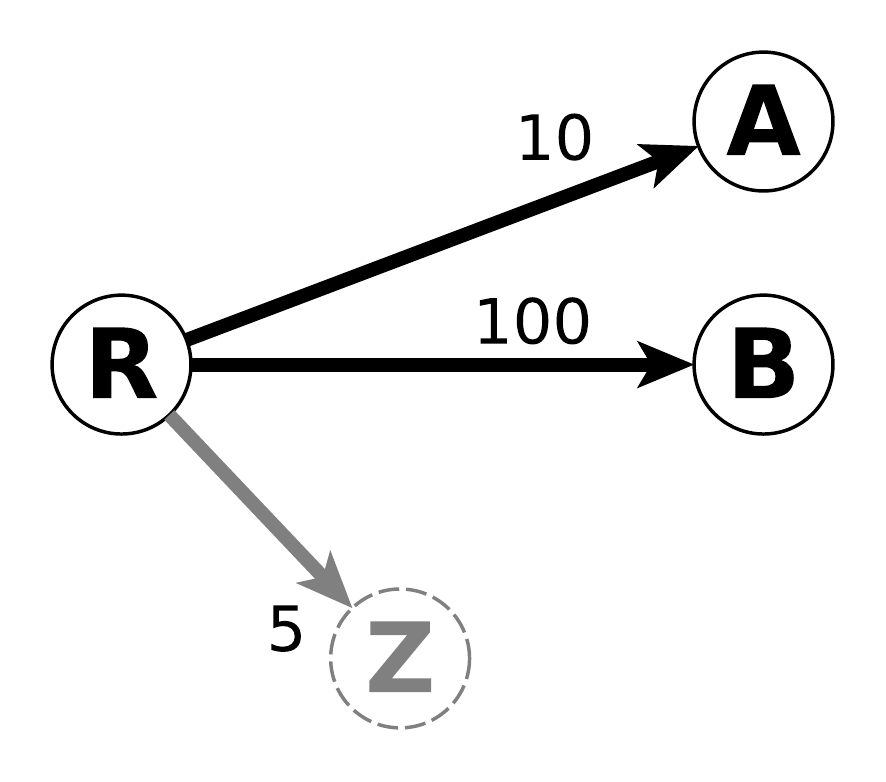}~~~}\hspace{.5cm}
    \subfloat[\scriptsize Anomalous pattern.\label{fig:forwardingExample1}]{~~~\includegraphics[width=.28\columnwidth]{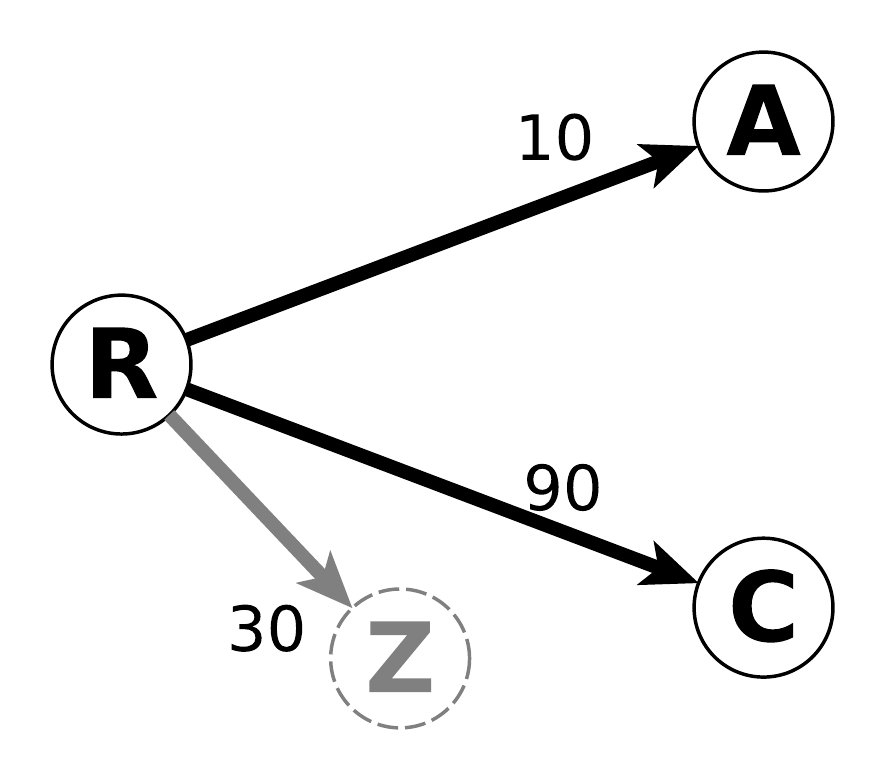}~}
    \vspace*{-2mm}
    \caption{Two forwarding patterns for router $R$. 
        $A,B$, and $C$ are next hops identified in traceroutes.
$Z$ shows packet loss and next hops that are unresponsive to traceroute.}
    \vspace*{-2mm}
\end{figure}

Figure \ref{fig:forwardingExample0} illustrates the example of a router $R$ with 
two responsive hops, $A$ and $B$, and unresponsive hop, $Z$.
The packet forwarding pattern of this router is formally defined as a vector where 
each element represents a next hop and the value of the element is the number of 
packets transmitted to that hop. 
For Figure \ref{fig:forwardingExample0} the forwarding pattern of $R$ 
is $F^R=[10, 100, 5]$.

To summarize router $R$'s usual patterns and to update this reference
with new patterns, we again employ exponential smoothing.
Let $F^{R}_{t}=\{p_i | i \in [1,n]\}$ be the forwarding pattern for router $R$ at time $t$ and
$\bar F^{R}_{t-1}=\{\bar p_i | i \in [1,n]\}$ be the reference computed at time $t-1$. 
These two vectors are sorted such as $p_i$ and $\bar p_i$ correspond to the same next hop $i$.
If the hop $i$ is unseen at time $t$ then $p_i=0$, similarly, 
if the hop $i$ is observed for the first time at time $t$ then $\bar p_i=0$. 
The reference $\bar F^{R}_{t-1}$ is updated with the new pattern $F^{R}_{t}$ as follows:
\begin{equation}
    \bar F^R_{t} = \alpha F^R_{t} + (1-\alpha) \bar F^R_{t-1}.
\end{equation}
As in Section \ref{sec:normalRef}, a small $\alpha$ value allows us to mitigate the impact 
of anomalous values.
The reference $\bar F^R_t$ represents the usual forwarding pattern for router $R$
and is the normal reference used for the anomaly detection method discussed in the 
next section.
A reference $\bar F^R_{t}$ is valid only for a certain destination IP address. 
In practice we compute a different reference for each traceroute target; thus,
several references are maintained for a single router.

\subsection{Forwarding anomaly detection}
\label{sec:forwardingAnomaly}
\subsubsection{Correlation analysis}
Detecting anomalous forwarding patterns consists of identifying patterns
$F$ that deviate from the computed normal reference $\bar F$.
In normal conditions we expect a router to forward packets as they did in past
observations.
In other words, we expect $F$ and $\bar F$ to be linearly correlated.
This linear dependence is easily measurable as the Pearson product-moment correlation 
coefficient of $F$ and $\bar F$, hereafter denoted as $\rho_{F,\bar F}$. 
The values of $\rho_{F,\bar F}$ range in $[-1,1]$.
Positive values mean that the forwarding patterns expressed by $F$ and $\bar F$ 
are compatible, while negative values indicate opposite patterns hence forwarding
anomalies.
Therefore, all patterns $F$ with a correlation coefficient $\rho_{F, \bar F}<\tau$
are reported as anomalous.
In our experiments we arbitrarily set $\tau=-0.25$, as the empirical distribution of $\rho_{F, \bar F}$
features a knee around that value.
Conservative results can be obtained with lower $\tau$ values, but higher values 
are best avoided as $\rho>-0.25$ represents very weak anti-correlation. 

\subsubsection{Anomalous next hop identification}
\label{sec:forwardingResponsibility}
When a forwarding pattern $F$ is reported as anomalous, it means that the proportions
of packets sent to next hops are different from those observed in the past.
Further, an anomalous pattern can be caused by just a few aberrant next hops.
We devise a metric to identify hops that are responsible for forwarding pattern changes.
Let $F~=~\{p_i|i \in [1,n]\}$ be an anomalous pattern and $\bar F = \{\bar p_i|i \in [1,n]\}$
the computed normal reference. 
Then we quantify the responsibility of the next hop $i$ to the pattern change as:
\begin{equation}
    r_i = -\rho_{F,\bar F} \frac{p_i - \bar p_i}{\sum_{j=1}^{n} | p_j - \bar p_j|}.
    \label{eq:responsibility}
\end{equation}
The responsibility metric $r_i$ ranges in $[-1,1]$. 
Values close to zero mean that the next hop $i$ received an usual number of packets
thus it is likely not responsible for the pattern change.
On the other hand, values deviating from 0 indicate anomalous next hops.
Positive values stand for hops that are newly observed, while negative values
represent hops with an unusually low number of packets.

For example, assume Figure \ref{fig:forwardingExample0} depicts $\bar F^R$, the computed 
normal reference for router $R$, and Figure \ref{fig:forwardingExample1} illustrates
$F^R$, the latest forwarding pattern observed.
The correlation coefficient for these patterns, $\rho_{F^R,\bar F^R}=-0.6$, is lower 
than the threshold $\tau$, thus $F^R$ is reported as anomalous. 
The responsibility scores for $A,B,C$, and $Z$ are, respectively, $0, -0.28, 0.25,$ 
and $0.07$;
suggesting that packets are ordinarily transmitted to $A$ and $Z$, but, the number 
of packets to $B$ is abnormally low while the count to $C$ is exceptionally high. 
In other words traffic usually forwarded to $B$ is now going through $C$.
In the case of a next hop dropping a significant number of packets, the responsibility 
score of this hop will be negative while the score of $Z$ will be large.

\section{Detection \TFIDF{and Characterization} of major events}
\label{sec:networkdisruptions}
The proposed delay analysis method (\autoref{sec:delayAnalysis}) and packet 
forwarding model (\autoref{sec:forwardingAnomalies}) are both
designed to report anomalies found in large-scale traceroute measurements.
With RIPE Atlas these methods allow us to monitor hundreds of thousands links 
and potentially obtain a large number of alarms (i.e. either delay changes or 
forwarding anomalies).
Investigating each alarm can be very tedious and time consuming.
In this section we introduce a simple technique to aggregate alarms and  
report only significant network disruptions. 

\noindent{\bf Alarm aggregation.} \
Major network disruptions are characterized by either a large-scale alteration of numerous
links or exceptional connectivity issues at a one or more locations. 
We wish to emphasize both by aggregating alarms based on their temporal and 
spatial characteristics.
The temporal grouping of alarms allows us to highlight large-scale events impacting many routers
at the same time.
Similarly, collecting alarms that are topologically close allows us to emphasize network 
disruptions bound to a particular entity.
In early experiments we have tried several spatial aggregations, including geographical ones, 
and found that grouping alarms per AS is relevant because 
most significant events are contained within one or a few ASs. 

Consequently, we group delay change alarms by the reported IP pair and 
forwarding anomalies by the next hops' IP addresses.
The IP to AS mapping is done using longest prefix match, and alarms with IP
addresses from different ASs are assigned to multiple groups.

Alarms from each AS are then processed to compute two time series representing 
the severity of reported anomalies, thus the AS's condition.
The severity of anomalies is measured differently for delay change and packet 
forwarding alarms.
For delay changes the severity is measured by the deviation from 
the normal reference, $d(\Delta)$ (Equation \ref{eq:delay:deviation}).
Severity of forwarding anomalies is given by $r_i$, the responsibility score of 
the reported next hop $i$ (Equation \ref{eq:responsibility}).
Thereby, AS network conditions are represented by two time series, one is the sum of $d(\Delta)$ 
over time and the other the sum of $r_i$ over time.
In the case of forwarding anomalies, $r_i$ values are negative if a hop from the 
AS is devalued and positive otherwise. 
Consequently, if traffic usually goes through a router $i$ but is suddenly rerouted to 
router $j$, and both $i$ and $j$ are assigned to the same AS, then the negative
$r_i$ and positive $r_j$ values cancel out, thus the anomaly was mitigated at the AS level.

\noindent{\bf Event detection.} \
Finding major network disruptions in an AS is done by identifying peaks
in either of the two time series described above.  We implement a simple
outlier detection mechanism to identify these peaks.

Let $X=\{x_t | t \in \mathbb{N}\}$ be a time series representing delay changes or
forwarding anomalies for a certain AS and $mag(X)$ be the 
magnitude of the AS network alteration defined as:
\begin{equation}
    mag(X) = \frac{X - \median(X)}{1+1.4826 \mad(X)} 
    \label{eq:mag}
\end{equation}
where $\median$ and $\mad$ are the one-week sliding median and median absolute
deviation
\cite{wilcox:stats10}.
In the following sections we report magnitude scores found with our dataset
and investigate corresponding network disruptions.

\section{Results}
\label{sec:results}
Using the Internet-wide traceroutes from RIPE Atlas (\autoref{sec:dataset}), we report 
delay changes and forwarding anomalies from eight months in 2015 and 1060 ASs. 
In the following we present aggregate results of the identified delay changes and forwarding anomalies.
Then, we dive into case studies showing the relevance of the proposed methods 
to detect and locate network disruptions of different types (\autoref{sec:kroot}, \ref{sec:results_tm}, and \ref{sec:results_amsix}).

\noindent{\bf Delay changes.} \
\label{sec:results_delay}
In our experiments we monitored delays for 262k IPv4 links (42k IPv6 links).
On average links are observed by 147 IPv4 probes (133 IPv6 probes)
and 33\% of the links were reported to have at least one abnormal delay change.

\begin{figure}[t]
    \vspace*{-3mm}
    \subfloat[\scriptsize Complementary cumulative distribution function for the delay change magnitude. Prominent changes are on the right hand side.\label{fig:overview_delay}]{\includegraphics[width=.48\columnwidth]{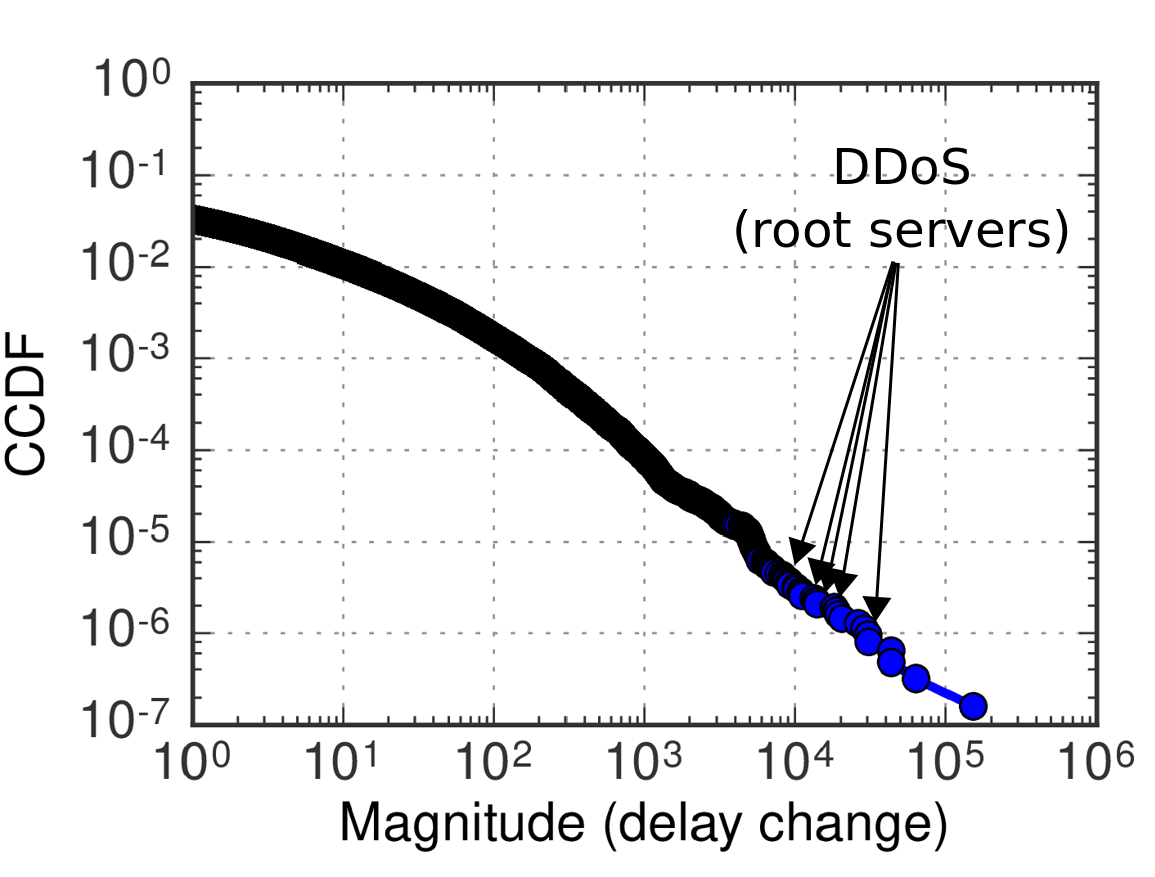}}\hfill
    \subfloat[\scriptsize Cumulative distribution function of the forwarding anomaly magnitude. Prominent anomalies are on the left hand side.\label{fig:overview_forwarding}]{\includegraphics[width=.48\columnwidth]{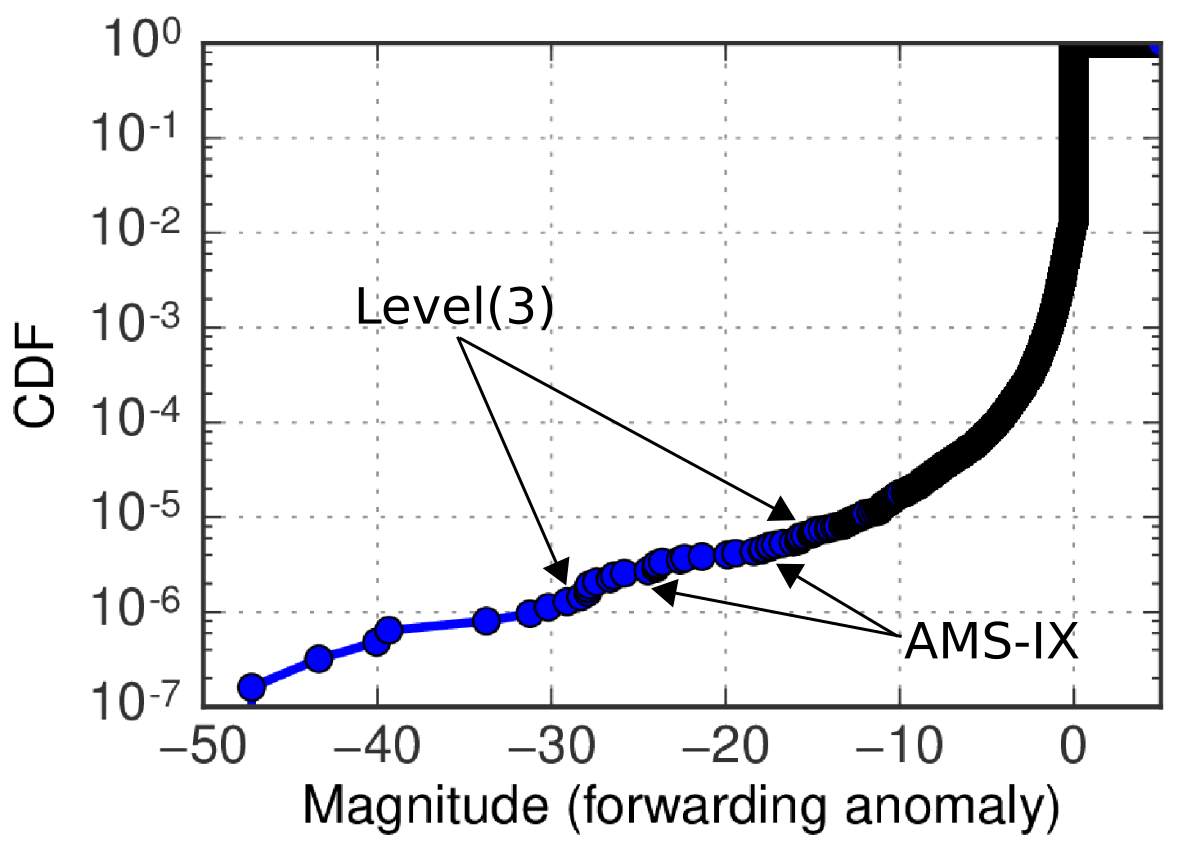}}
    \vspace*{-3mm}
\caption{Distribution of hourly magnitude for all ASs. Arrows point to prominent anomalies presented in the three study cases.}
\end{figure}

We computed the hourly delay change magnitude for each monitored ASs, 
Figure \ref{fig:overview_delay} depicts the distribution of all these values.
97\% of the time we observe a magnitude lower than 1, meaning that ASs are usually free of large transient delay changes. 
The heavy tail of the distribution, however, indicates that delay changes can have a very detrimental impact on Internet delays.
We manually inspected the most prominent delay changes but found that validating such results is particularly hard as public reports are rarely available and Internet service providers are reluctant to disclose troubles that occurred in their networks.
In Section \ref{sec:kroot}, we detail a DDoS attack that generated congestion in several ASs and accounts for 5 of the top 23 delay changes reported in our dataset (Fig.~\ref{fig:overview_delay}).

Furthermore, in accordance with the central limit theorem, we observe a narrower 
confidence interval for links visited by numerous probes; hence a better differential
RTT estimation and the ability to detect smaller delay changes.

\noindent{\bf Forwarding anomalies.} \
\label{sec:results_forwarding}
Using RIPE Atlas traceroutes, we also computed packet forwarding models for 170k IPv4 router IPs (87k IPv6 router IPs).
These are the number of router IP addresses found in traceroutes; to resolve these to routers IP alias resolution techniques should be deployed \cite{keys:ton13}.
On average forwarding models contain four different next hops over the eight months of data. 

We computed the hourly forwarding anomaly magnitude for each AS, Figure \ref{fig:overview_forwarding} illustrates the distribution of these values.
This distribution features a heavy left tail representing a few significant forwarding anomalies due to important packet loss or traffic redirection.
Namely, forwarding anomaly magnitude is lower than $-10$ for only 0.001\% of the time.
Similarly to the delay changes, validating these results is challenging.
In Section \ref{sec:results_tm} and \ref{sec:results_amsix} we investigate two significant events from the top 20 forwarding anomalies found in our dataset (Fig.~\ref{fig:overview_forwarding}). 
These events are already publicly documented but the proposed method provides 
further insights on their location and impact.

\subsection{DDoS attack on DNS root servers}
\label{sec:kroot}
Our first case-study shows the impact of a large distributed denial-of-service 
(DDoS) attack on network infrastructure.
The simplest form of DDoS attack consists of sending a huge number of requests to 
a targeted service, overwhelming the service and leaving little or no 
resources for legitimate use.
The extremely large amount of traffic generated by this type of attack is not only detrimental
to the victim but also routers in its proximity.

\begin{figure}
    \includegraphics[width=.9\columnwidth]{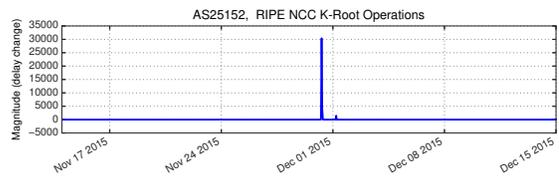}
    \caption{Delay change magnitude for AS25152 reveals the two DDoS against
    the K-root server.}
    \label{fig:krootMagnitude}
    \vspace*{-5mm}
\end{figure}
\begin{figure*}[t]
    \captionsetup[subfigure]{labelformat=empty}
    \captionsetup*[subfigure]{position=bottom}
    \subfloat[\label{fig:kroot:kansas}]{\raisebox{20mm}{(a)}\includegraphics[width=.45\columnwidth]{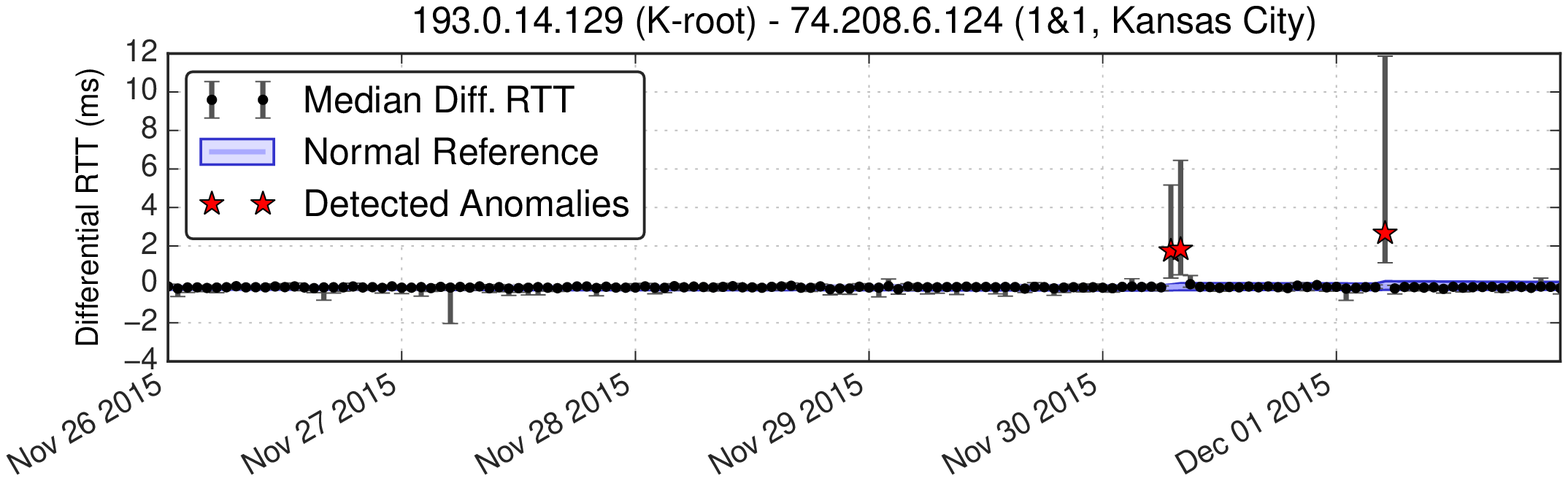}}~~~~~~~~~~
    \subfloat[\label{fig:kroot:pl}]{\raisebox{20mm}{(b)}\includegraphics[width=.45\columnwidth]{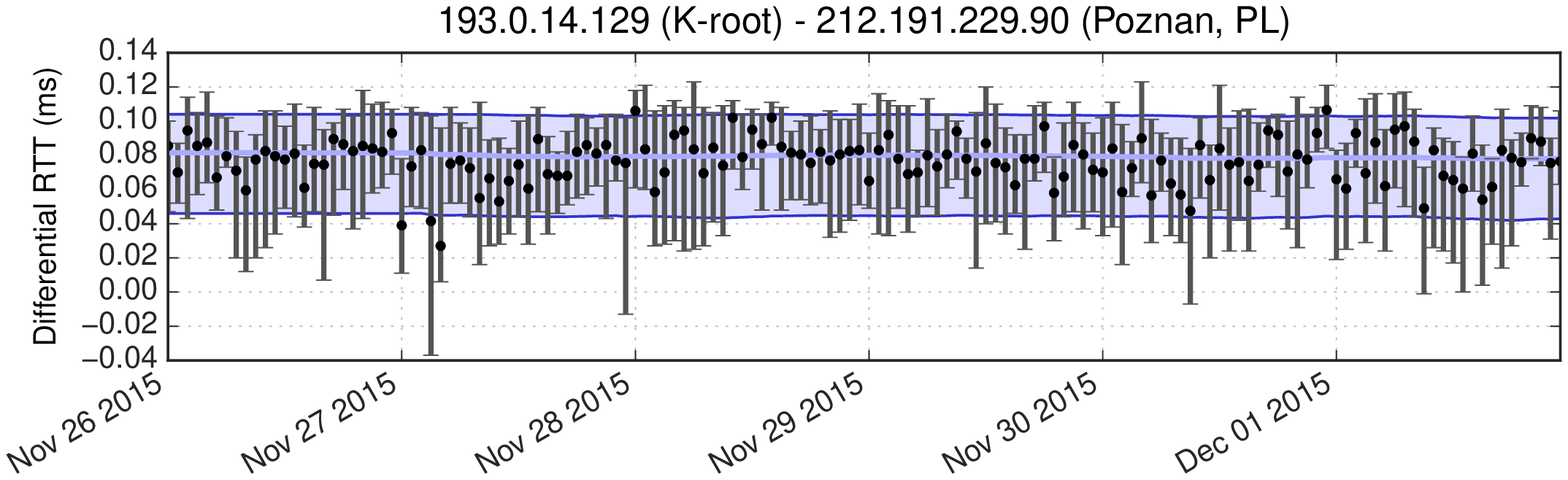}}\\
    \vspace*{-8mm}
    \subfloat[\label{fig:kroot:decix0}]{\raisebox{20mm}{(c)}\includegraphics[width=.45\columnwidth]{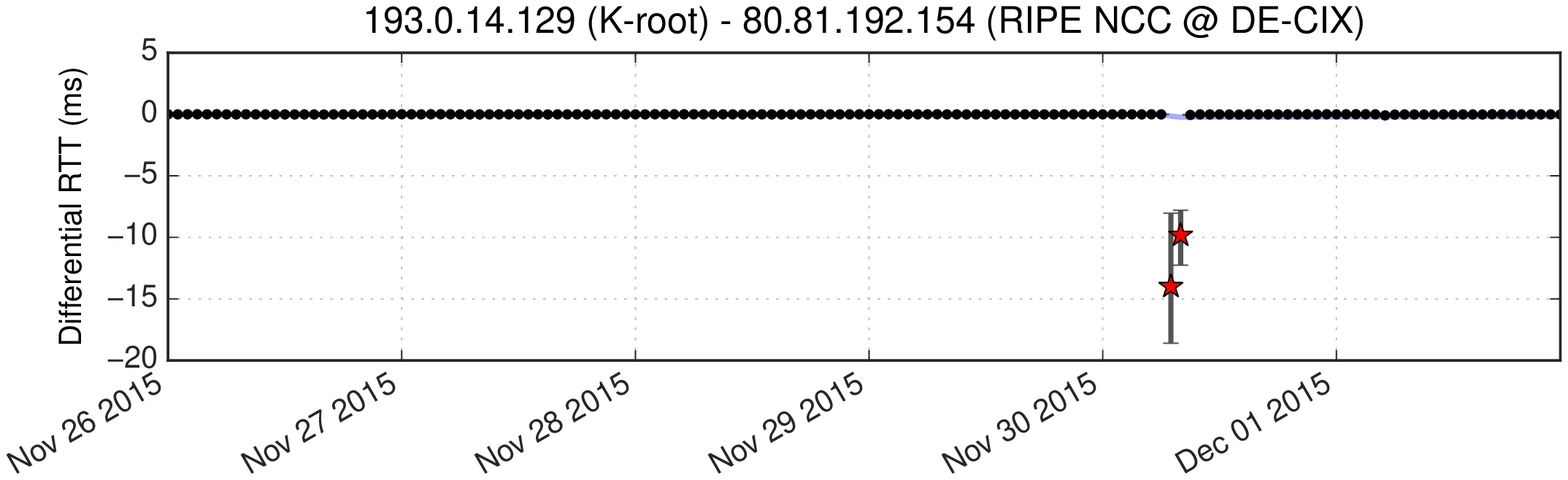}}~~~~~~~~~~
    \subfloat[\label{fig:kroot:stpe0}]{\raisebox{20mm}{(d)}\includegraphics[width=.45\columnwidth]{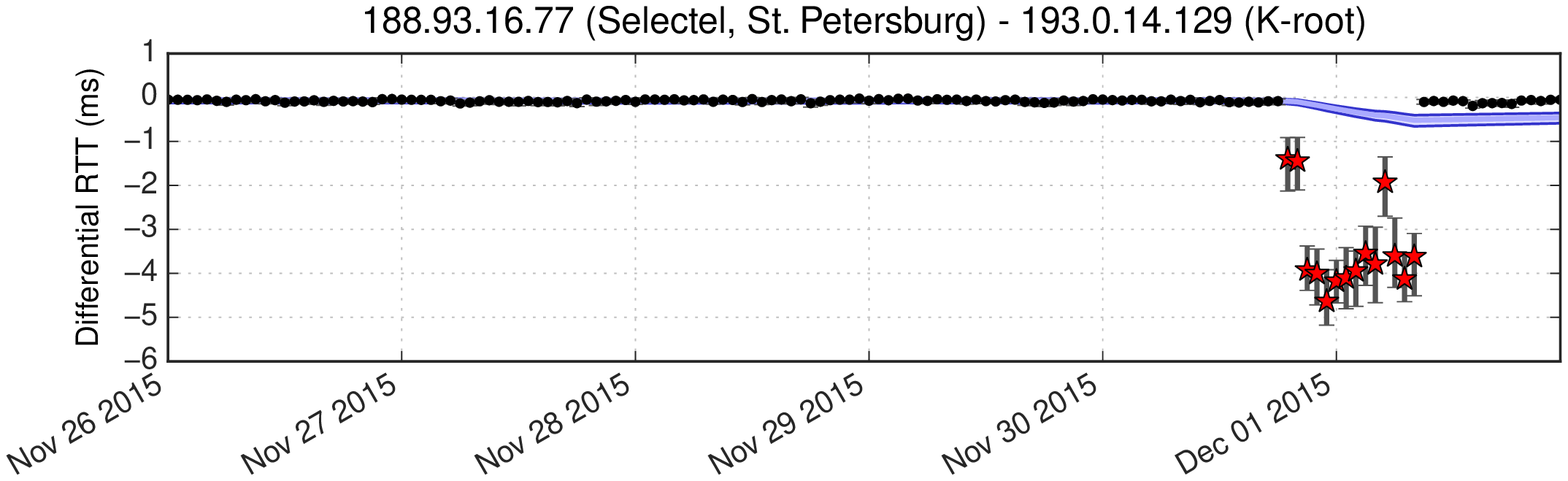}}\\ 
    \vspace*{-8mm}
    \subfloat[\label{fig:kroot:decix1}]{\raisebox{20mm}{(e)}\includegraphics[width=.45\columnwidth]{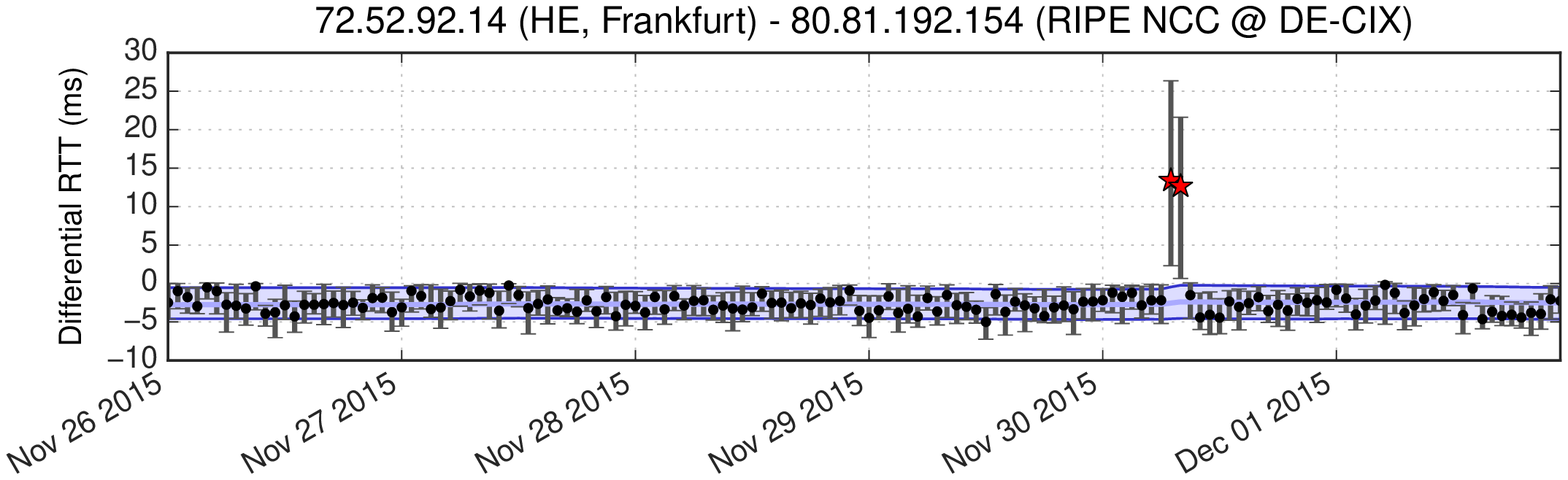}}~~~~~~~~~~
    \subfloat[\label{fig:kroot:stpe1}]{\raisebox{20mm}{(f)}\includegraphics[width=.45\columnwidth]{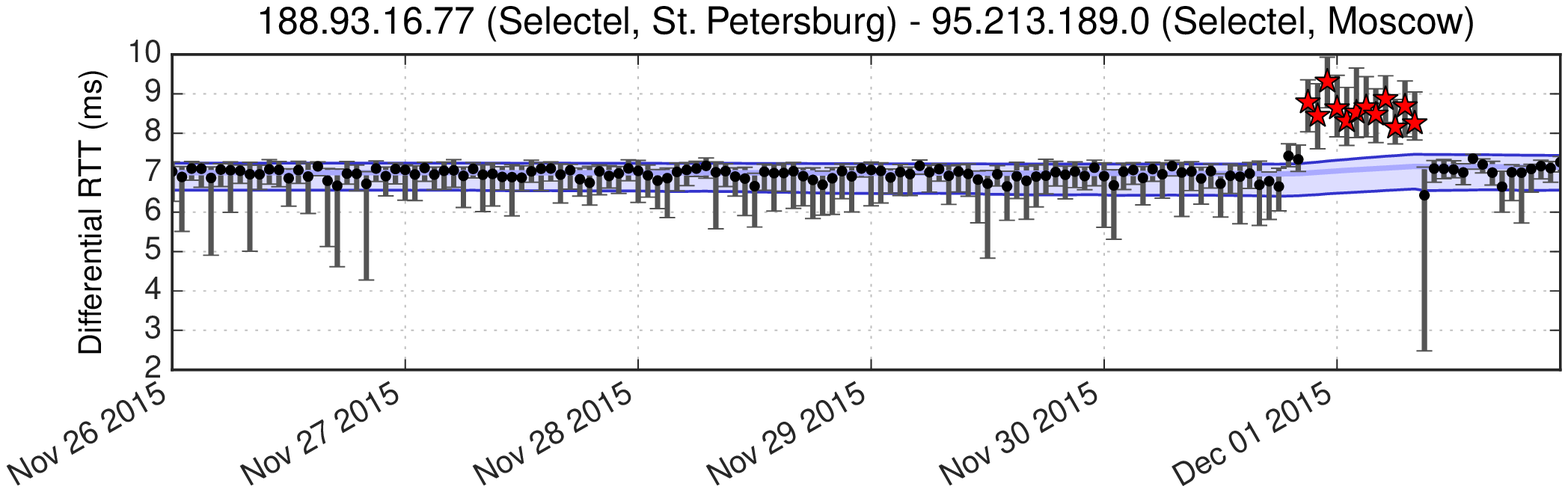}} \\
    \caption{Examples of delay change alarms reported during the DDoS attacks against
    DNS root servers. The attacks have differently impacted the connectivity of K-root server instances.} 
    \vspace*{-8mm}
\end{figure*}
\begin{figure}
    \includegraphics[width=\textwidth]{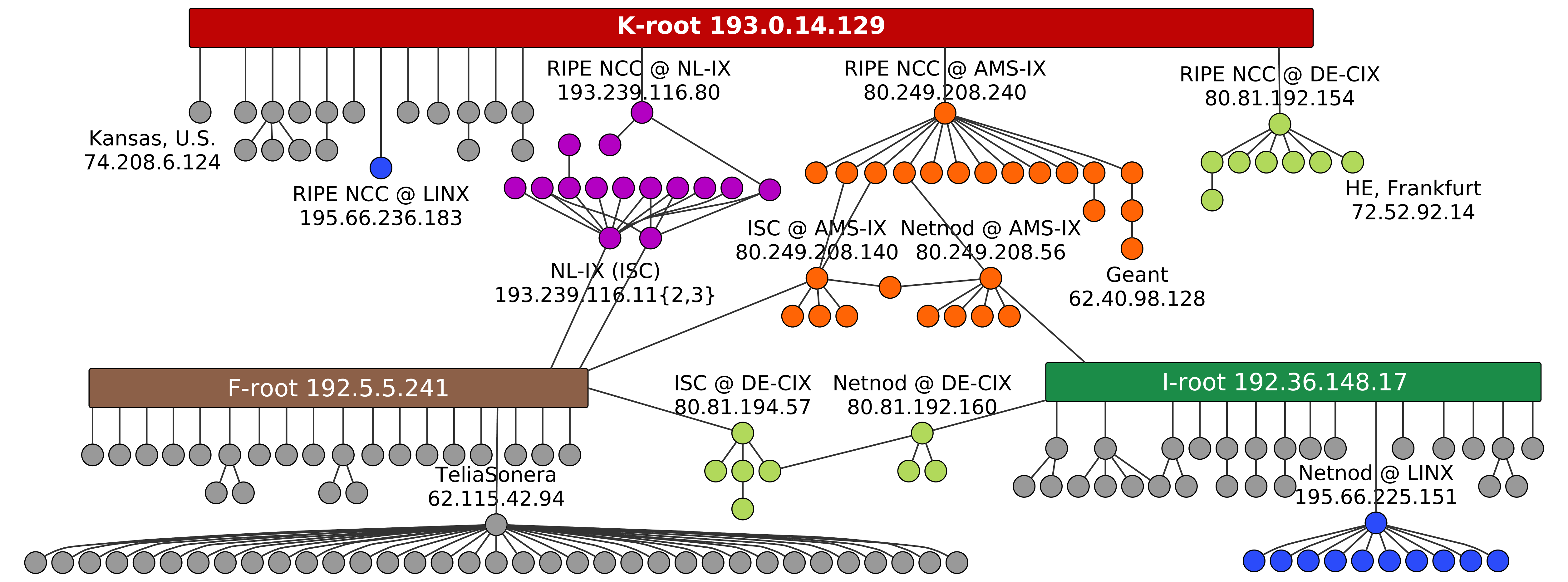}
    \caption{Alarms reported on November \nth{30} at 08:00 UTC and related to the 
    K-root server. Each node represent an IPv4 address, edges stand for reported alarms.
    Rectangular nodes represent anycast addresses, hence distributed infrastructure. 
    Circular node colors represent IP addresses related to certain IXPs.\label{fig:kroot:graph}}
\end{figure}

We investigate network disruptions caused by two DDoS attacks against
DNS root servers.  These attacks have been briefly documented by root
server operators \cite{root:ddos15,verisign:oarc24}.  The first attack
was on November \nth{30} from 06:50 to 09:30 UTC, the second on December
\nth{1} from 05:10 until 06:10 UTC.  As the source IP addresses for both
attacks were spoofed, it is unclear from reports \cite{verisign:oarc24}
where the traffic originated.

Thanks to the K-root operators, we were able to
carefully validate our results for the attack toward the K name 
server and the corresponding AS (AS25152).

\noindent{\bf Event detection.} \
Monitoring the delay change magnitude for AS25152 clearly shows 
the two attacks against the K-root infrastructure (Fig.~\ref{fig:krootMagnitude}).
The two peaks on November \nth{30} and December \nth{1} highlight important disruptions
of an unprecedented level. 
The first peak spans from 07:00 to 09:00 UTC and the second from 05:00 to 06:00 UTC,
which correspond to the intervals reported by many server operators.
\TFIDF{
Further, the $\tfidf$ score of the K-root IP address is the highest meaning 
that this IP address is the most reported during the two events thus
confirming the nature of the event.
}

The highest forwarding anomaly magnitude for AS25152 is recorded on 
November \nth{30} at 08:00 and is negative ($mag(X)=-0.5$), meaning that only a few packets 
have been dropped in ASs hosting root servers.
These observations match the server operators' reports and emphasize the 
strength of anycast in mitigating such attacks.

\noindent{\bf In-depth analysis: K-root.} \
A key advantage of our method is reporting delay changes per link, 
allowing us to precisely locate the effects of the two attacks in the network.
Reported delay changes contain one IP address for each end of the link. 
Delay changes detected on the last hop to the K-root server are identified by the server
IP address (193.0.14.129) and the router in front of it.
Since K-root is anycast, the actual location 
of a reported server instance must be revealed by locating the adjacent router.
For example, Figure \ref{fig:kroot:kansas} depicts the differential RTT for an IP
pair composed of the K-root IP address and a router located in Kansas City; hence
this link represents the last hop to the K-root instance in Kansas City.

During the two attacks we saw alarms from 23 unique IP pairs containing 
the K-root server address.
Different instances were impacted differently by the attacks.
First, we found instances affected by both attacks, for example the one in Kansas
City (Fig.~\ref{fig:kroot:kansas}) is reported during the entire period of time
documented by server operators.
Second, we also observed instances impacted by only one attack, see Figure \ref{fig:kroot:decix0}.
The most reported instance during that period is the one deployed in 
St.~Petersburg (Fig.~\ref{fig:kroot:stpe0}).
For this instance abnormal delays are observed for 14 consecutive hours. 
A possible explanation for this is that hosts topologically close to this instance
caused anomalous network conditions for a longer period of time than other reported
DDoS intervals.
Finally, thanks to anycast, for some instances we did not record anomalous network
conditions.
Figure \ref{fig:kroot:pl} illustrates the differential RTT for an instance in Poland
that exhibits very stable delays. 
The corresponding normal reference is exceptionally narrow and constant even during the attacks.

Not only are the last hops to K-root instances  detected by our method;
we also observe other links with important delay changes.
Figure \ref{fig:kroot:decix1} depicts a link in the Deutscher Commercial Internet Exchange
(DE-CIX) which is upstream of the K-root instance in Frankfurt (Fig.~\ref{fig:kroot:decix0}).
This link between Hurricane-Electric (AS6939) and the K-root AS
exhibits a 15ms delay change (difference between the median differential
RTT and the reference median) during the first attack.
The upstream link of the instance in St.Petersburg (Fig.~\ref{fig:kroot:stpe1})
is also significantly altered during the attack and is consistent with the peculiar 
changes observed for this instance (Fig.~\ref{fig:kroot:stpe0}).
In certain cases, we observed effects of the attack even further upstream.
For example, we observe 7.5ms delay change on a link in the Geant network three 
hops away from the K-root server (see \emph{Geant 62.40.98.128} in Fig.~\ref{fig:kroot:graph}).

To assess the extent of the attacks on the network, we create a graph, where
nodes are IP addresses and links are alarms generated from differential RTTs between these IP addresses.
Starting from the K-root server, we see alarms with common IP addresses, and 
obtain a connected component of all alarms connected to the K-root server.
Figure \ref{fig:kroot:graph} depicts the connected component involving K-root for delay changes detected on 
November \nth{30} at 08:00 UTC.
An anycast address is illustrated by a large rectangular node,
because it represents several physical systems. 
Figure \ref{fig:kroot:graph} does not show the physical topology 
of the network but a logical IP view of reported alarms.
Each edge to an anycast address usually represents a different instance of a root
server.
There are rare cases where two edges may represent the same instance, for example,
the K-root instance available at AMS-IX and NL-IX is actually the same physical cluster.
Some of the alarms mentioned above and illustrated in Figure \ref{fig:kroot:kansas}, \ref{fig:kroot:decix0},
and \ref{fig:kroot:decix1} are also displayed in Figure \ref{fig:kroot:graph}.
The shape of the graph reveals the wide impact of the attack on network infrastructure.
It also shows that alarms reported for the K-root servers are adjacent to the ones
reported for the F and I-root servers.
This is due to the presence of all three servers at the same exchange points;
hence some network devices are affected by malicious traffic targeting multiple
 root servers.
The concentration of root servers is of course delicate in this situation.
Although packet loss at root servers has been negligible, we found significant
forwarding anomalies at their upstream providers.
For example, AMS-IX (AS1200) shows a forwarding anomaly magnitude of $-24$ during that incident.

Additional root servers are represented by different connected components.
During the three hours of attack there were 129 alarms involving root servers for IPv4 (49 for IPv6).
In agreement with the observations made by servers operators \cite{verisign:oarc24},
we observed no significant delay change for root servers A, D, G, L, and M.

\begin{figure}[t]
    \includegraphics[width=.9\columnwidth]{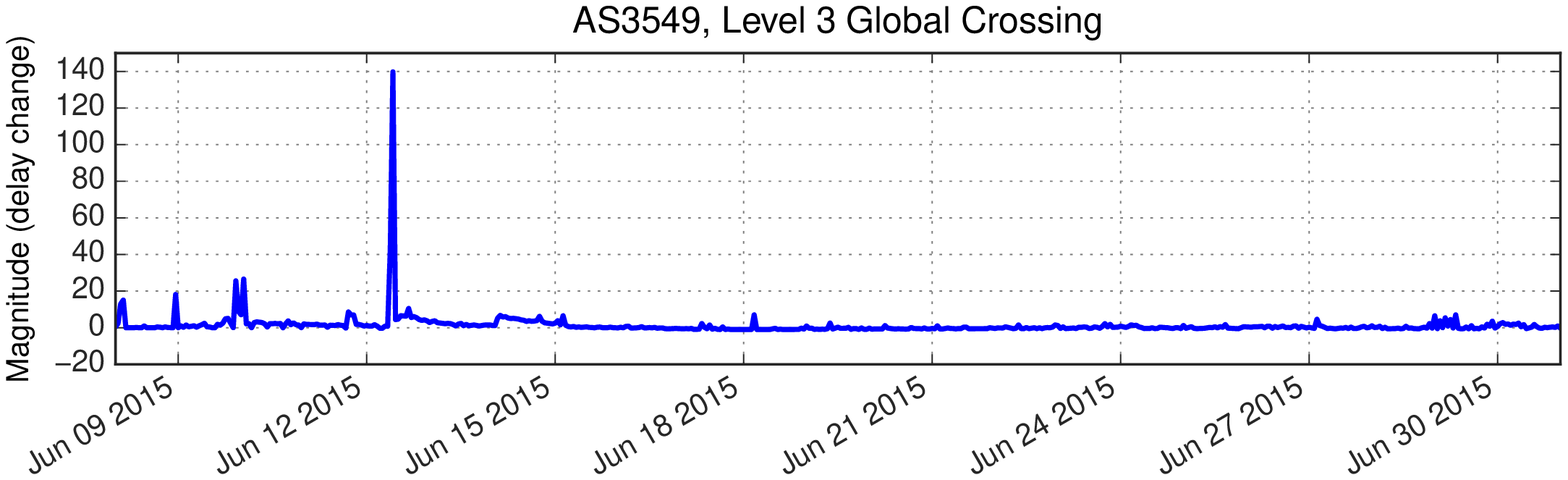}\\
    \includegraphics[width=.9\columnwidth]{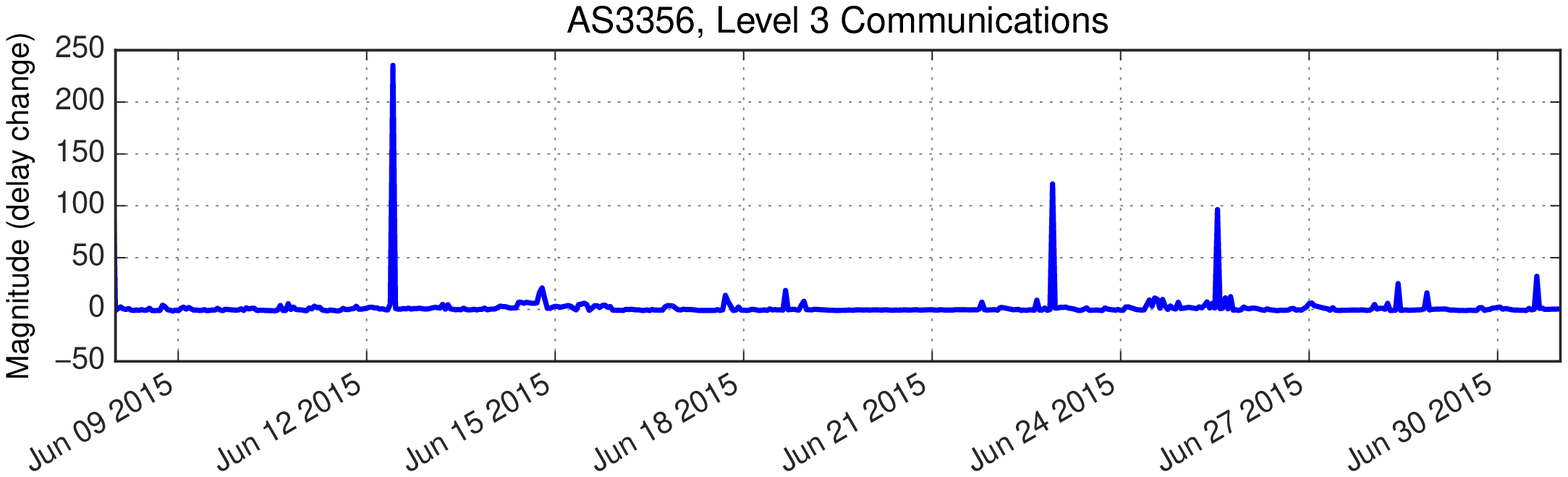}
    \vspace*{-8mm}
    \caption{Delay change magnitude for all monitored IP addresses in two Level(3) ASs.}
    \label{fig:l3:delay}
\end{figure}
\begin{figure}[t]
    \includegraphics[width=.9\columnwidth]{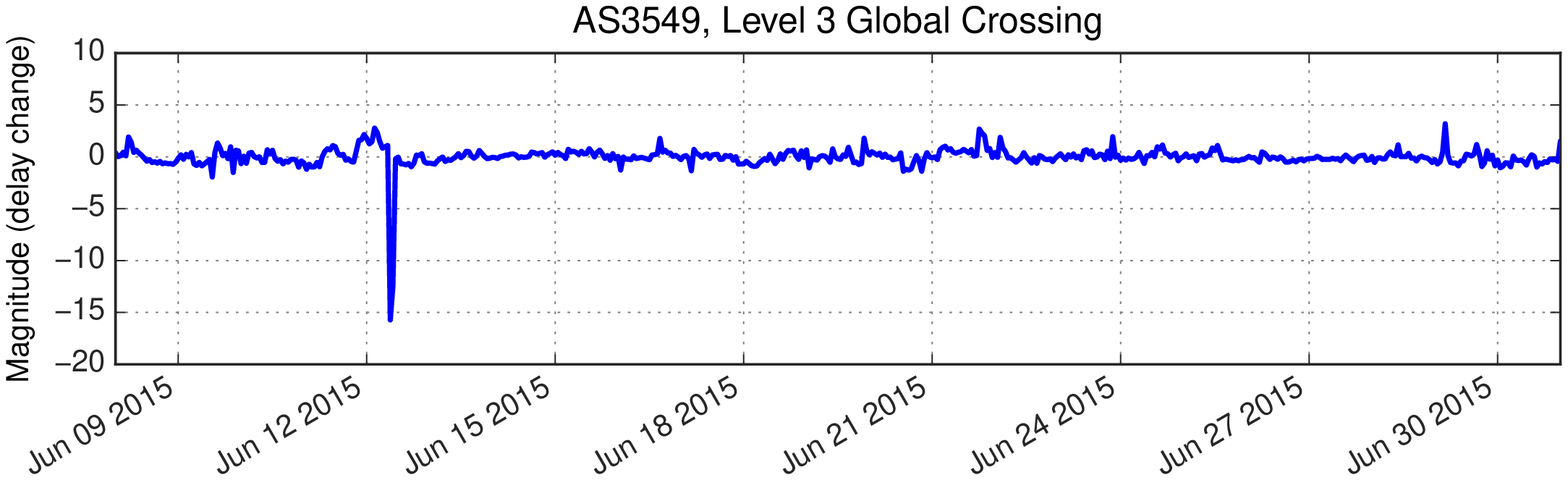}\\
    \includegraphics[width=.9\columnwidth]{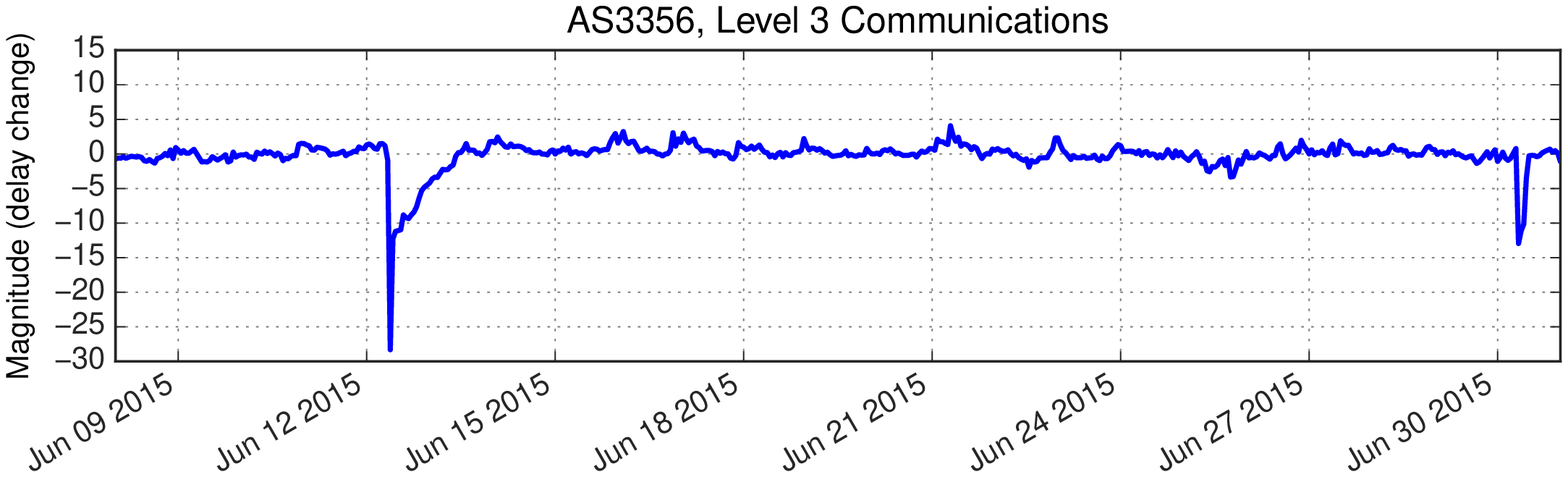}
    \vspace*{-8mm}
    \caption{Forwarding anomaly magnitude for all monitored IP addresses in two Level(3)ASs.}
    \label{fig:l3:forwarding}
\end{figure}

\begin{figure}[t]
    \subfloat[\scriptsize London-London link: delay change reported on June~\nth{12} at 09:00 and 10:00 UTC.\label{fig:l3:ex0}]{\includegraphics[width=.9\columnwidth]{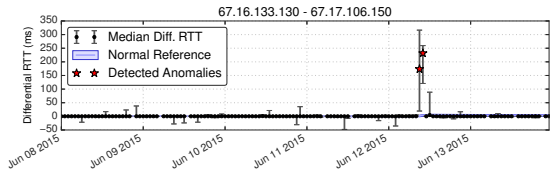}}\\
    \subfloat[\scriptsize New York-London link: delay change reported at 10:00 UTC. RTT samples 
    for June \nth{12} at 09:00 UTC are missing due to forwarding anomaly (packet loss).\label{fig:l3:ex1}]{\includegraphics[width=.9\columnwidth]{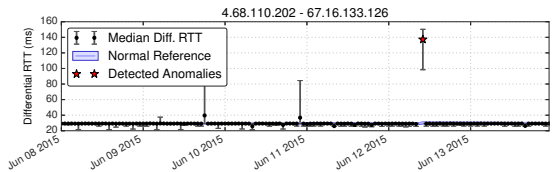}}
    \vspace*{-3mm}
    \caption{Example of delay change alarms reported during the Telekom Malaysia 
    BGP route leak for two links from Level3 networks.}
    \label{fig:l3:example}
\end{figure}

\begin{figure}
    \includegraphics[width=.8\columnwidth]{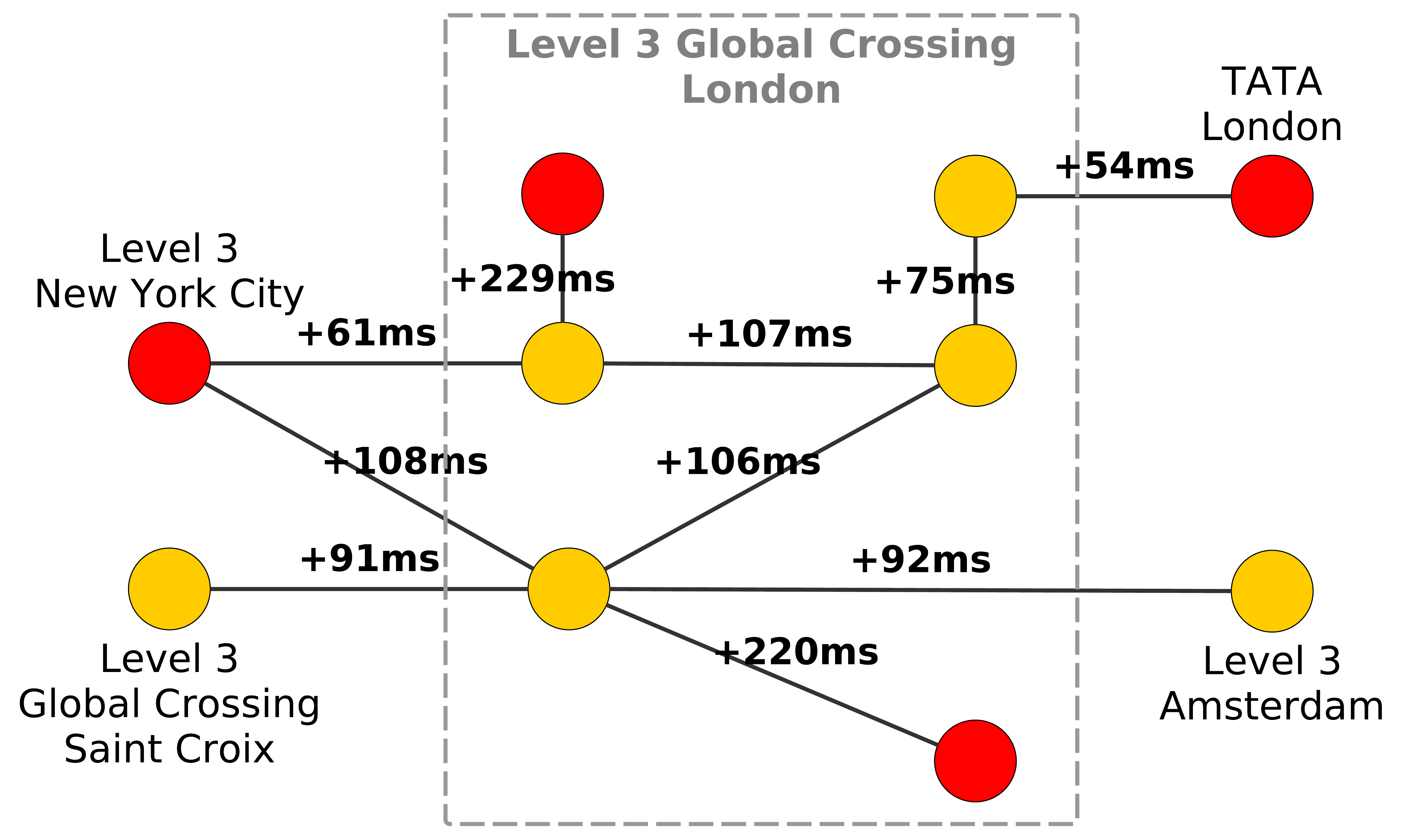}
    \caption{Congestion at Level(3) Global Crossing (AS3549) in London on June \nth{12} 2015.
    Each node represents an IPv4 address, edges represent delay changes for an IP pair.
    Red nodes depict IP addresses involved in forwarding anomalies.}
    \label{fig:l3:graph}
\end{figure}

\subsection{Telekom Malaysia BGP route leak}
\label{sec:results_tm}
The above example of the K-root servers illustrates the benefits of our
delay change detection method in detecting anomalies near a small AS at the edge.
In this section we investigate network disruptions for a tier 1 ISP showing
that the methods also enable us to monitor large ASs containing numerous links.
This case study also exposes a different type of network disruption; here the 
detected anomalies are caused by abnormal traffic rerouting.

On June \nth{12} 2015, 08:43 UTC, Telekom Malaysia (AS4788) unintentionally sent BGP announcements 
for numerous IP prefixes to its provider Level(3) Global Crossing (AS3549) which accepted
them. 
The resulting traffic attraction to Telekom Malaysia caused
latency increases for Internet users all over the globe.
The event was acknowledged by Telekom Malaysia \cite{tmnet:leak15},
and independently reported by BGP monitoring projects \cite{bgpmon:tmnet,1000eyes:tmnet}.
Connectivity issues have been mainly attributed to congested peering links
between Telekom Malaysia and Level(3) Global Crossing. 
In the remainder of this section we investigate the impact of rerouted traffic on Level(3)
Global Crossing (AS3549) and its parent company, Level(3) Communications (AS3356),
showing worldwide disruption.

\noindent{\bf Network disruptions in Level(3).} \
Monitoring delay changes and forwarding anomalies for the numerous links
that constitute the two Level(3) ASs is made easy with the magnitude metric. 
Figure \ref{fig:l3:delay} and \ref{fig:l3:forwarding} depict the magnitude in terms of,
respectively, delay change and forwarding anomaly for the two Level(3) ASs in June 2015.
The two positive peaks in Fig. \ref{fig:l3:delay} and the two negative peaks in Fig. \ref{fig:l3:forwarding}
are all reported on June \nth{12} from 09:00 to 11:00 UTC, exposing the impact of rerouting on both ASs.
The overall delay increased for both ASs, but AS3549 was most affected.
The negative forwarding anomaly magnitudes (Fig.~\ref{fig:l3:forwarding})
show that routers from both ASs were disappearing abnormally from the forwarding 
model obtained by traceroute.
At the same time packet loss increased, implying that numerous routers from both
ASs dropped a lot of packets.
These are the most significant forwarding anomalies monitored for Level(3) in our
8-month dataset.

\TFIDF{
The event characterization method indicates that two subnets are responsible for 
96\% of the delay deviation observed for AS3549 (4.68.110.0/24 and 4.69.167.0/24),
and one subnet is responsible for 73\% of the delay deviation for AS3356 (67.16.133.0/24).
Forwarding anomalies involve a larger number of IP addresses, especially during the
first hour following the BGP route leaks.
}

\noindent{\bf In-depth analysis.} \
Reverse DNS lookups of reported IP addresses suggests congestion was seen
in numerous cities, including, Amsterdam, Berlin, Dublin, Frankfurt,
London, Los Angeles, Miami, New York, Paris, Vienna, and Washington, for both Level(3) ASs.
Figure \ref{fig:l3:example} shows the differential RTT obtained for two links
located in New York and London.
Both links exhibit significant delay increases synchronous with the Telekom Malaysia route leak.
The London-London link (Fig.~\ref{fig:l3:ex0}) is reported from 09:00 to 11:00 UTC,
while the New York-London link (Fig.~\ref{fig:l3:ex1}) is reported from 10:00 to 11:00 UTC.
The IP address identified in New York is found in forwarding anomalies, and is suspected
of dropping probing packets from 09:00 to 10:00 UTC; hence preventing the collection 
of RTT samples for this link.
This example illustrates the complementarity of the delay change and forwarding 
anomaly detection methods.

As in the case of the K-root servers, several adjacent links are reported at the same
time. 
Figure \ref{fig:l3:graph} shows related components of alarms reported on June \nth{12}
at 10:00 UTC in London.
The label on each edge is the absolute difference between the observed median differential RTT 
and the median of the normal reference.
The links in Fig. \ref{fig:l3:ex0} and \ref{fig:l3:ex1} are marked by
delay changes of, respectively, \emph{+229ms} and \emph{+108ms}.
Similar observations are made for the two Level(3) ASs and numerous cities mainly in U.S. and Europe.
Consequently, even non-rerouted traffic going through Level(3) at that time could also incur 
significant latency increase and packet loss. 


\subsection{Amsterdam Internet Exchange Outage}
\label{sec:results_amsix}
The first two study cases presented network disruptions with significant delay changes.
Here we introduce an example of network disruption visible only
through forwarding anomalies; showing the need for both delay change and forwarding
anomaly detection methods.
In this example the disruption is caused by a technical fault in an Internet exchange
resulting in extensive connectivity issues. 

On May \nth{13} 2015 around 10:20 UTC, the Amsterdam Internet Exchange (AMS-IX) encountered
substantial connectivity problems due to a technical issue during maintenance activities. 
Consequently, several connected networks could not exchange traffic through the 
AMS-IX platform; hence a number of Internet services were unavailable \cite{amsix:outage15}.
AMS-IX reported that the problem was solved at 10:30 UTC; but traffic statistics
indicate that the level of transmitted traffic did not return to normal until 12:00 UTC \cite{kisteleki:amsix15,emile:amsix15}.

\begin{figure}
    \includegraphics[width=.9\columnwidth,clip]{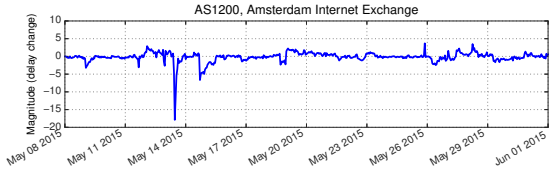}
    \caption{Forwarding anomaly magnitude for the Amsterdam Internet Exchange 
    peering LAN (AS1200).}
    \vspace*{-3mm}
    \label{fig:amsix}
\end{figure}

\noindent{\bf Event detection.} \
The delay change method did not conclusively detect this outage, due to
lack of RTT samples
during the outage. 
Indeed, the packet loss rate showed significant disturbances at AMS-IX. 
These changes were captured by our packet forwarding model as a sudden 
disappearance of the AMS-IX peering LAN for many neighboring routers.
Consequently, forwarding anomalies with negative responsibility scores 
(Equation \ref{eq:responsibility}) were synchronously reported for IP addresses in
the AMS-IX peering LAN.
Monitoring the magnitude for the corresponding AS (Fig.~\ref{fig:amsix}) 
reveals these changes as a significant negative peak on May \nth{13} 11:00 UTC. 
Further, the coincidental surge of unresponsive hops reported by forwarding anomalies
supports the fact that traffic was not rerouted but dropped.
The packet forwarding model allows us to precisely
determine peers that could not exchange traffic during the outage.
In total 770 IP pairs related to the AMS-IX peering LAN  
became unresponsive.
Therefore, the proposed method to learn packet forwarding patterns and systematically identify
unresponsive IP addresses greatly eases the understanding of such an outage. 


\section{Internet Health Report}
\label{sec:ihr}

The key contribution of our method is to allow operators to 
troubleshoot connectivity issues outside
their own network, normally a very difficult task.
Typical circumstances include distant users of other ISPs complaining
that an ISP's web service is unavailable, or local customers complaining
to their ISP about connectivity issues, though their ISP's network is
not the cause of the issues.
In these cases being able to pinpoint the exact location of the problem
allows operators to contact the appropriate NOC, or to consider routing
decisions to avoid unreliable networks. 


In order to provide a practical tool to network operators, we have 
integrated the proposed methods with the RIPE Atlas streaming API.
This gives us near-real time traceroutes for all long-lived Atlas measurements (including built-in and anchoring measurements) and enables us to detect events in a timely manner.
Our results are publicly available through an interactive website \cite{ihr:website} and an
API \cite{ihr:api} such that researchers and operators can access
computed results in an easy and systematic way.  Of course, an operator
can take our code and run it against the Atlas streaming API themselves,
focusing on only the part(s) of the topology which interests them \cite{ihr:code}.
Thanks to an increasing number of Atlas measurements and probes, the number of monitored ASs is constantly increasing. 
As of April 2017, we were monitoring a total of 5,436 ASs, a significant fraction of the 7,800 transit ASs observed in the Internet \cite{geoff:bgp2016}. 

We encourage operators interested in using our system to deploy Atlas anchors in their network so that probes will automatically initiate traceroute towards their network, and visited transit links will be monitored by our system.
The results  enable operators to easily monitor the diverse transit
networks between their infrastructure and the thousands of Atlas probes
deployed world-wide.




\section{Conclusions}
\label{sec:conclusion}
In this paper we investigated the challenges to monitoring network conditions
using traceroute results.
We then tackled these challenges with a statistical approach that took advantage
of large-scale traceroute measurements to accurately pinpoint delay
changes and forwarding anomalies.
Our experiments with the RIPE Atlas platform validated our methods and
emphasized the benefits of this approach to characterize topological impacts. 

The methods proposed in this paper complement the literature by circumventing 
common problems found in past work. 
With the help of the packet forwarding model, we take advantage of all collected 
traceroutes including even those that are incomplete due to packet loss.
Also, as we do not rely on any IP or ICMP options, the number of monitored routers 
is superior to previous work.
In fact, our statistical approach allows us to study any link with
routers responding
to traceroute and that can be seen by probes hosted in at least three different ASs.
Therefore, the number of monitored links mainly depends on the placement of probes and 
the selected traceroute destinations.
In other words, using our techniques the number of monitored links is given 
by the measurement setup rather than the router's implementation.
Stub ASs hosting probes but no traceroute targets were not monitored as they were
observed only by probes from the same AS.
In the case of symmetric links we could release the probe diversity constraint. 
However, due to the current lack of efficient technique to assert an arbitrary link symmetry we leave this task for future work .

We make our tools and results publicly available 
\cite{ihr:website,ihr:api,ihr:code}
in order to share our findings and contribute
to a better understanding of Internet reliability. 

\newpage

%
\bibliographystyle{abbrv}

\bibliography{references}  

\arxiv{
\appendix
\section{Qualitative Comparison}
\label{sec:comparison}
As opposed to tulip \cite{mahajan:sosp03}, cing \cite{cing:infocom03}, Pong \cite{deng:icnp08} and TSLP \cite{luckie:imc14}, the main benefits of our proposal are its compliance with current router functionalities, its robust statistical analysis and the recycling of existing data.

All techniques, including ours, require routers sending back ICMP packets for TTL-expired packets.
In addition, TSLP requires routers to implement IP options (pre-specified timestamps or record route). 
Tulip and cing require routers to implement ICMP Timestamp and have 
strong assumptions for IP ID implementation. 


In addition to these restrictions, some techniques have coverage limits. 
Pong can only monitor paths between probes, TSLP considers only inter-domain symmetric links adjacent to the probes' ASs, and our system is constrained to links monitored by probes from at least 3 different ASs.

Finally, tulip and cing are consuming significantly more network resources than other methods. 
These two methods rely on ICMP timestamps and require a large number of samples to correct routers' clocks artifacts. 
Consequently, the authors of tulip estimate delays on a path using 1000 measurements per router plus an extra 500 measurements per router for packet loss estimation \cite{mahajan:sosp03}.
In contrast, our system requires as little as nine packets per router and is designed to take advantage of existing traceroute data, thus really adding no extra load to the network.

\section{Theoretical Limitations}
\label{appendix:theo_limitations}

The sensitivity of our approach ino detecting abnormal delay changes depends 
mainly on the size of the time bin which is based on probes deployment and probing rate.
A link is monitored only if it is traversed from vantage points within at least three 
different ASs (Section \ref{sec:probeDiversity}). 
As traceroute sends three packets per hop, for a link we expect at least $m=3*3$ 
packets per time bin.  
Consequently,  the number of vantage points monitoring a link and their 
probing rate $r$ (i.e. number of traceroutes per hour) determine the minimum usable 
time bin $T_{min}=\frac{m}{3rn}$.
Intuitively experiments with many probes or a high probing rate would permit the 
use of short time bin.

Let $T \geq T_{min}$ be the selected time bin, then $3rnT$ is the expected number 
of packets obtained for a link per time bin.
Because our approach relies on the median, 50\% of these packets should be
impacted by an event to be detected.
In other words, an event is detected if it affects more than $1+\frac{3rnT}{2}$
packets within a time bin.
Consequently, the smallest detectable event in hour is:
\begin{equation}
    \frac{1}{3rn}(1+\frac{3rnT}{2})=\frac{1}{3rn}+\frac{T}{2}.
    \label{minEvent}
\end{equation}

In Section \ref{sec:results} we analyze builtin measurements which initiate 
traceroute every 30 minutes ($r=2$ traceroutes per hour) thus the minimum usable 
time bin is $T_{min}=0.5$ hour.
In our experiments we conservatively set the time bin $T=1$ hour, hence, according
to Equation \ref{minEvent}, the shortest event we can detect for a link monitored 
by three vantage points ($n=3$) is 33 minutes.
Because of the higher probing rate of anchoring measurements ($r=4$), one could 
detect events lasting only nine minutes with this dataset.


}
\end{document}